\documentclass[prd,nofootinbib,a4paper,showpacs,superscriptaddress,floatfix]{revtex4-1}
\usepackage{amsmath}
\usepackage{amssymb}
\usepackage{graphicx}
\usepackage{hyperref}

\newcommand{\tr}{\mathop{\mathrm{tr}}}

\newcommand{\V}[1]{{\boldsymbol{#1}}}

\begin{document}
\def\Regensburg{Institute for Theoretical Physics, University of Regensburg, 93040 Regensburg, Germany}

\author{Gunnar S. Bali}
\affiliation{\Regensburg}
\affiliation{Tata Institute of Fundamental Research, Homi Bhabha Road, Mumbai 400005, India}

\author{Sara Collins}
\affiliation{\Regensburg}

\author{Stephan D\"urr}
\affiliation{Bergische Universit\"at Wuppertal,
   Gau{\ss}stra{\ss}e\,20, 42119 Wuppertal, Germany}
\affiliation{J\"ulich Supercomputing Center,
   Forschungszentrum J\"ulich, 52425 J\"ulich, Germany}

\author{Issaku Kanamori}
\email{issaku.kanamori@physik.uni-regensburg.de}
\affiliation{\Regensburg}
\affiliation{
Institute of Physics,
National Chiao-Tung University,
Hsinchu 30010,
Taiwan}

\date{\today}
\pacs{12.38.Gc, 13.20.Fc, 14.40.Lb, 14.40.Be}
\title{$D_s \rightarrow \eta, \eta'$ semileptonic decay form factors
with disconnected quark loop contributions}

\begin{abstract}
We calculate for the first time the form factors of the semileptonic
decays of the $D_s$ meson to $\eta$ and $\eta^\prime$ using lattice
techniques. As a by-product of the calculation we obtain the masses
and leading distribution amplitudes of the $\eta$ and $\eta^\prime$ mesons. 
We use $N_f=2+1$
non-perturbatively improved clover fermions on configurations with
a lattice spacing $a\sim 0.075$~fm. 
We are able to obtain clear signals for relevant matrix elements,
using several noise reduction techniques, both for the connected and
disconnected contributions. This includes a new method for reducing the
variance of pseudoscalar disconnected two-point functions.
At zero momentum transfer, we obtain for the scalar form factors,
$|f_0^{D_s\to \eta}|=0.564(11)$ 
and
$|f_0^{D_s\to \eta'}|=0.437(18)$
at $M_\pi\approx 470\, {\rm MeV}$,
as well as
$|f_0^{D_s\to \eta}|=0.542(13)$ 
and
$|f_0^{D_s\to \eta'}|=0.404(25)$
at $M_\pi\approx 370\, {\rm MeV}$,
where the errors are statistical only.
\end{abstract}

\maketitle

\section{Introduction}

In general, semileptonic decays of charmed mesons are well studied
both experimentally and theoretically, in particular, using lattice
techniques. However, this is not the case for the $D_s$ meson for
which the main semileptonic modes are to the $\phi$, $\eta$ and
$\eta^\prime$ mesons.  Lattice studies of these decays are technically
challenging due to the presence of disconnected quark-line
contributions. So far only the form factor for the decay $D_s\to\phi
\ell\bar{\nu}_{\ell}$ has been computed, omitting the disconnected
contributions~\cite{Donald:2013pea}.  QCD sum rules provide an
alternative approach based on the operator product expansion (OPE) and
analyticity using the assumption of quark-hadron duality.  There is
one result using the standard local OPE in terms of condensates
\cite{Colangelo:2001cv} and two using light cone OPE in terms of
distribution amplitudes \cite{Azizi:2010zj,Offen:2013nma}.  These
studies utilize the $\eta$ and $\eta^\prime$ distribution amplitudes,
which, in principle, can be calculated on the lattice. A first
principles calculation of the form factors for $D_s\to\eta^{(\prime)}
\ell\bar{\nu}_{\ell}$ therefore can serve as a cross-check on the
assumptions of the sum rule approach and is of phenomenological
interest in itself, providing information on the internal structure of
the mesons in the final state (see, for example,
Ref.~\cite{DiDonato:2011kr}).  In terms of experimental results, there
are no measurements of the form factors for these modes so far and only the
branching fractions for $D_s \to\eta^{(\prime)} \ell\bar{\nu}_{\ell}$
have been determined by the CLEO collaboration
\cite{Yelton:2009aa}.

In this article, we report on our exploratory study of
the $D_s$ to $\eta^{(\prime)} \ell \bar{\nu}_{\ell}$ semileptonic decay
form factors using lattice techniques.
Some preliminary results have been presented in
Refs.~\cite{Kanamori:2013rha, Collins:2013lxa, Collins:2014qsa}.
The relevant matrix elements for these decay modes are parameterized
as follows
\begin{equation}
 \langle \eta^{(\prime)}(k)| V_\mu |D_s(p)\rangle
 = f_+(q^2) \left[ (p+k)_\mu - \frac{M_{D_s}^2 - M_{\eta^{(\prime)}}^2}{q^2}q_\mu\right]
   + {f_0(q^2)}  \frac{M_{D_s}^2 - M_{\eta^{(\prime)}}^2}{q^2}q_\mu\,,
 \label{eq:vector-current}
\end{equation}
where $V_\mu$ is a vector current at position\footnote{In practice
one averages 
$V_{\mu}(\V{y})e^{i\V{q}\cdot\V{y}}$ 
over all positions $\V{y}$, injecting the spatial
momentum $\V{q}$ required by momentum conservation, to increase
statistics.} $\mathbf{0}$, $q_\mu = p_\mu - k_\mu$ is the four-momentum transfer and $M_{D_s}$ and $M_{\eta^{(\prime)}}$ are
the masses of the $D_s$ and the $\eta^{(\prime)}$ mesons, respectively.
This matrix element is characterized by two form factors, 
$f_0(q^2)$ and $f_+(q^2)$.
In this work we focus on the scalar form factor $f_0(q^2)$,
which we can also obtain from a scalar current $S=\bar{s} c$ \cite{Na:2010uf}:
\begin{equation}
 f_0(q^2) 
= \frac{m_c - m_s}{M_{D_s}^2 - M_{\eta^{(\prime)}}^2}\langle \eta^{(\prime)} |S| D_s \rangle.
 \label{eq:scalar-current}
\end{equation}
We use this relation because the combination $(m_c-m_s) S$ 
(and therefore $f_0(q^2)$) is a renormalization group invariant, provided
the vector mass difference 
$m_c-m_s=\left(\kappa_c^{-1}-\kappa_s^{-1}\right)/(2a)$
is used. Eq.~(\ref{eq:scalar-current}) is also free of additive
renormalization.

\begin{figure}[bp]
$\displaystyle
\raisebox{-2em}{\includegraphics[width=15em]{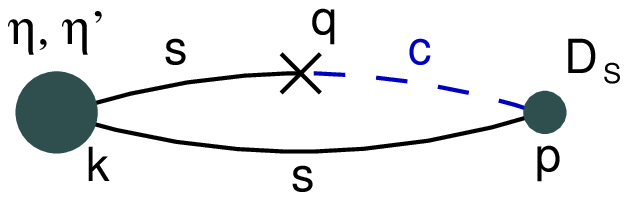}}
-\sum_{l=u,d,s}\left(
\raisebox{-2em}{\includegraphics[width=15em]{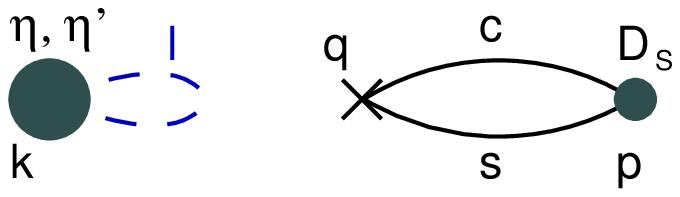}}
\right)
$

\caption{Connected (first term) and disconnected (second term) 
fermion loop diagrams. We use stochastic methods to calculate the blue
dashed fermion lines.
The labeling of
the four-momenta $p,q,k$ reflects the conventions adopted in 
Eq.~(\ref{eq:vector-current}).
}
\label{fig:3pt}
\end{figure}

The three-point function needed to compute the form factor contains
quark-line disconnected loops (see Fig.~\ref{fig:3pt}). Often corrections
from disconnected loops are numerically small. However,
their impact on the three-point function is enhanced
by a factor of about three, due to the summation over three
light quark flavours. Moreover, in the pseudoscalar case
the disconnected quark loops couple to the axial anomaly. 
In spite of the computational expense and the
inferior quality of the signal, relative to that of the
quark-line connected contribution, the calculation
of the disconnected contribution
turns out to be feasible and its impact is significant
\cite{Bali:2011yx, Kanamori:2013rha}. 
Therefore,
these decay modes also provide a perfect playground
for testing a variety of techniques for calculating the disconnected
quark-line loops.

We use QCDSF $N_f=2+1$ configurations 
\cite{Bietenholz:2010jr, Bietenholz:2011qq} that
were generated using a novel approach for varying the sea quark masses,
which is ideal for studying flavour physics in the SU(3) flavour basis.
The flavour singlet mass average of the three light quarks,
 $\frac{1}{3}(m_u+m_d+m_s)$, is kept fixed so that
the combination $2M_K^2 + M_\pi^2$ computed from
the kaon mass, $M_K$, and pion mass, $M_\pi$,
approximately  coincides with the physical value.
Starting from the flavour SU(3) symmetric point ($m_u=m_d=m_s$), 
$m_l=m_u=m_d$ is reduced as $m_s$ is increased.

The outline of this paper is as follows: in the next Section we
describe the technical details of the lattice calculation.  Before we
can address decays of the $D_s$ meson into final states including the
$\eta$ or $\eta'$ mesons, we have to construct the corresponding
interpolators. Therefore, in Section \ref{sec:eta-etap-states}, we
determine the mixing of the physical states relative to the
octet-singlet basis. We present a new method to reduce statistical
noise and obtain the $\eta$ and $\eta'$ masses and the leading
distribution amplitudes.  The details of the new method, described in
Subsection \ref{sec:finite_volume}, are quite technical and can be
skipped by those readers who are primarily interested in the final
results. In Section~\ref{sec:form factor} we describe our methods for
extracting the matrix elements relevant for the computation of the
form factors. Subsequently, these are obtained in the same Section,
before we conclude.

\section{Details of the lattice calculation}

The QCDSF $N_f=2+1$ configurations
were generated with the tree level Symanzik improved gluon action
and the Stout Link Non-perturbatively improved Clover 
fermion action (SLiNC) \cite{Cundy:2009yy}. 
We use the same action for the valence-only charm quark.
The SLiNC action is on-shell $O(a)$ improved.
In general, there will be $O(a)$ correction
term, $a\bar{s}\stackrel{\leftrightarrow}{D}_{\mu}\!\!\gamma_{\mu}c$,
to Eq.~(\ref{eq:scalar-current}).
However, this term can be eliminated using the equations of motion
and one can show that the non-singlet improvement
coefficients $b_S=-2b_m$~\cite{Sint:1997dj,Bhattacharya:2005rb}
cancel from
Eq.~(\ref{eq:scalar-current}) so that
$f_0(q^2)$ is automatically $O(a)$-improved.

The parameters are summarized in Table~\ref{tab:configs}.
So far we have only used one lattice spacing $a\sim 0.075\, {\rm fm}$
(determined using the quantity $w_0$ proposed in Ref.~\cite{Borsanyi:2012zs}),
and one volume $V_4=L^3T=24^3\times 48\,a^4$,
which corresponds to a physical spatial extent $L\sim 1.8\, {\rm fm}$.
Our value of the lattice spacing is about 10\% smaller than 
the value of Refs.~\cite{Bietenholz:2010jr, Bietenholz:2011qq}
($a\sim 0.083\,{\rm fm}$) which was obtained from the average
octet baryon mass, but is consistent with a newer determination 
($a\sim 0.074(2)\,{\rm fm}$) in Ref.~\cite{Horsley:2013wqa}.
We analyzed 939 configurations at the flavour symmetric
point ($m_l=m_s$), for which $M_\pi=M_K=470.5(1.8)\, {\rm MeV}$ (Set S),
and 239 configurations
($m_{l}< m_s$) with $M_\pi=370.1(3.3)\, {\rm MeV}$ and 
$M_K=509.1(2.7)\,{\rm MeV}$ (Set A).  
Due to the different value for the lattice spacing, these masses differ from
the numbers given in Refs.~\cite{Bietenholz:2010jr, Bietenholz:2011qq}.
In particular, the average octet pion mass exceeds the experimental
value $[(M_\pi^2+2M_K^2)/3]^{1/2}\approx 411\,\mathrm{MeV}$ by
about $60\,\mathrm{MeV}$,
meaning that extrapolating to the physical pion mass, we
would end up with unphysically heavy kaons.
The charm quark mass $m_c$ was tuned so that the spin averaged
1S charmonium mass, 
$M_{\overline{1S}}=\frac{1}{4}(M_{\eta_c} + 3 M_{J/\psi})$,
corresponds to the experimental value \cite{Bali:2011dc}.

In order to reduce autocorrelations, 
the configurations were sampled every 5 Monte Carlo trajectories for Set S
and every 10 trajectories for Set A.  
In addition, the location of the source was chosen randomly 
on each configuration.
However, significant correlations were found in the data  when calculating
the masses of the $\pi$, $\eta$ and $\eta'$ mesons,
and we chose a conservative bin size of 5 (25 molecular dynamics time
units) for Set S and 2 (20 molecular dynamics time units)
for Set A.
The mass of the $D_s$, the mixing angle discussed in
Sec.~\ref{sec:mixing}
and the form factor, $f_0(q^2)$, did not show 
any significant autocorrelations,
so we did not use binning for these observables.

For all source and sink interpolators,
we used a gauge invariant Gaussian smearing 
(Wuppertal smearing \cite{Gusken:1989ad,Gusken:1989qx}) 
with APE smeared gauge fields \cite{Falcioni:1984ei} in the spatial directions.
The smearing parameters were chosen to minimize excited state contributions
to the connected two-point functions.

\begin{table}
\caption{The simulation parameters.
Set S corresponds to the SU(3) flavour symmetric point where the pion,
kaon, and eta mesons are mass-degenerate while in Set A the symmetry between
the strange quark and the light quarks is broken.
The lattice size is $24^3 \times 48$ in both cases and $\beta=10/g^2$,
rather than $\beta=6/g^2$.}
\label{tab:configs}
\begin{ruledtabular}
  \begin{tabular}{ccccccc}
 Set & $\beta$ & $\kappa_l$ & $\kappa_s$ & $M_\pi$  & $L M_\pi $ & confs.\\
 \hline
  S  & 5.5     & 0.12090    & 0.12090    & $470.5(1.8)\,{\rm MeV}$ & 4.3 & 939 \\
  A  & 5.5     & 0.12104    & 0.12062    & $370.1(3.3)\,{\rm MeV}$ & 3.3 & 239
 \end{tabular}
\end{ruledtabular} 
\end{table}

Disconnected fermion loops appear in both two- and three-point functions.
These loops need to be evaluated at different times and momenta.
They can be obtained from the inverse of the dimensionless lattice 
Dirac operator, $M$, in the following way:
\begin{equation}
 C_{\rm 1pt}(t,\V{p}; \V{x}_0)
 = \sum_{\V{x}}\exp(i\V{p}\cdot(\V{x}-\V{x}_0)) \tr \left[
     \sum_{\V{x}',\, \V{x}''}
    \Gamma \phi(\V{x},\V{x}'') M^{-1}(t,\V{x}''; t,\V{x}')
    \phi(\V{x}',\V{x})
     \right]\, ,
 \label{eq:1pt-def}
\end{equation}
where the Dirac matrix for a pseudoscalar meson is $\Gamma=\gamma_5$ 
and $\phi$ is the smearing function.
The origin for the Fourier transformation is denoted as $\V{x}_0$.
Since $M$ satisfies $\gamma_5$-Hermiticity, the smearing function 
is Hermitian and commutes with $\gamma_5$,
the disconnected loop is real in coordinate space and 
$C_{\rm 1pt}(t,\V{p}) = C_{\rm 1pt}(t,-\V{p})^*$.
Details of the estimation of the disconnected loops
are given in Appendix \ref{app:disconn_loop}.

\section{$\eta$ and $\eta'$ states}
\label{sec:eta-etap-states}

Prior to determining decays of the $D_s$ into the $\eta$ or
$\eta'$ mesons, we have to construct these physical states.
We first discuss the general mixing formalism, relative
to the octet-singlet basis, then determine the respective
masses and compare our results to other studies. 
In Section \ref{sec:finite_volume}, 
which is technical and can be skipped on first reading, we discuss
a particular problem we encountered, due to the insufficient
sampling across topological sectors on one of our ensembles.
The method we suggest to resolve this turns out to be of a more
general applicability and significantly reduces statistical errors.
Finally, in Section \ref{sec:mixing} we determine mixing angles and leading 
distribution amplitudes of these states.

\subsection{Extracting physical states}
\label{sec:extracting_physical_states}

The correct creation operators for the $\eta$ and $\eta'$
states are a priori unknown in the flavour non-symmetric case (Set A).
We start from singlet 
$\eta_1 = \frac{1}{\sqrt{3}}(u\bar{u} + d\bar{d} + s\bar{s})$
and octet
$\eta_8 = \frac{1}{\sqrt{6}}(u\bar{u} + d\bar{d} -2 s\bar{s})$
states and
first calculate a $2\times 2$ correlation matrix of two-point
functions\footnote{We always use $\langle \cdot \rangle$ for
expectation values so that the correlation functions without
$\langle \cdot \rangle$ like $C_{\rm 1pt}(\V{p}, t)$ and 
$C_{\rm 2pt}(\V{p},t)$ denote configuration by configuration quantities.}
\begin{equation}
 \langle C_{\rm 2pt}(t,\V{p}) \rangle
=
\begin{pmatrix}
 \langle C_{\rm 2pt}^{88} (t,\V{p}) \rangle &
 \langle C_{\rm 2pt}^{81} (t,\V{p}) \rangle \\
 \langle C_{\rm 2pt}^{18} (t,\V{p}) \rangle &
 \langle C_{\rm 2pt}^{11} (t,\V{p}) \rangle
\end{pmatrix}
 \equiv
 \begin{pmatrix}
  \langle \mathcal{O}_8(t;\V{p}) \mathcal{O}_8^\dagger(0) \rangle
&   \langle \mathcal{O}_8(t;\V{p}) \mathcal{O}_1^\dagger(0) \rangle 
\\
  \langle \mathcal{O}_1(t;\V{p}) \mathcal{O}_8^\dagger(0) \rangle
&   \langle \mathcal{O}_1(t;\V{p}) \mathcal{O}_1^\dagger(0) \rangle 
 \end{pmatrix}\,,
\end{equation}
where $\mathcal{O}_8$ and $\mathcal{O}_1$ are smeared interpolators
for the octet and singlet states, respectively.
Each element includes disconnected fermion loop
contributions. The latter were averaged over all possible source
positions $x_0$ in space and time, shifting the source and the sink
accordingly.
For the connected part, we used low mode averaging 
\cite{DeGrand:2004qw,Giusti:2004yp}.
We describe the details of these calculations in Appendix~\ref{app:2pt}.
We solve the following generalized eigenvalue problem
\begin{align}
 \left\langle C_{\rm 2pt}(t_0,\V{p})\right\rangle^{-\frac{1}{2}}
\langle C_{\rm 2pt}(t,\V{p})\rangle
 v_{\alpha}(t,\V{p})
&=
 \lambda_{\alpha}(t,\V{p})
 \left\langle C_{\rm 2pt}(t_0,\V{p})\right\rangle^{\frac{1}{2}} 
  v_{\alpha}(t,\V{p})\,,
\end{align}
where $\lambda_{\alpha}(t,\V{p})$ ($\alpha=\eta, \eta'$) is the generalized
eigenvalue and $v_{\alpha}(t,\V{p})$ is the generalized eigenvector.
The time slice $t_0$ can be varied to minimize the excited state contributions
to $\lambda_{\alpha}$ and $v_{\alpha}$.
We tried $t_0/a=1$--$3$ and found no significant difference in the results, 
so we use $t_0/a=1$ which gives the largest range of $t>t_0$.
We parameterize the eigenvectors of the two-dimensional system
in the following way 
\begin{align}
 v_{\eta}(t, \V{p})&=(\cos\theta(t, \V{p}), -\sin\theta(t, \V{p}))^T\,,
&
 v_{\eta'}(t, \V{p})&=(\sin \theta'(t, \V{p}), \cos\theta'(t,\V{p}))^T\,.
 \label{eq:GEVeigenvectors}
\end{align}
Note that in general $\theta\neq\theta'$.
In the large $t$ limit, the ground state dominates,
$v_{\eta^{(\prime)}}(t,\V{p})\to v_{\eta^{(\prime)}}(\V{p})$ and
we can obtain the interpolators for the physical ground states:
\begin{align}
 \mathcal{O}_\eta 
  &= \cos\theta(\V{p})\, \mathcal{O}_8 - \sin\theta(\V{p})\, \mathcal{O}_1\,,
 &
 \mathcal{O}_{\eta'} 
 &= \sin\theta'(\V{p}) \, \mathcal{O}_8 + \cos\theta'(\V{p})\, \mathcal{O}_1\,.
 \label{eq:physical_interpolator}
\end{align}
It is sufficient to extract $\sin \theta$ and $\sin \theta'$.
This was done by fitting the corresponding components of 
$v_{\alpha}(t,\V{p})$ to a constant, taking into account correlations
including those between $\sin\theta$ and $\sin\theta'$ and those
between different time slices.
Using Eq.~(\ref{eq:physical_interpolator}),
we can construct the two-point functions of the physical states
for each $\V{p}$:
\begin{align}
 \langle C_{\rm 2pt}^{\eta}(t,\V{p}) \rangle
  &= \langle \mathcal{O}_\eta(t;\V{p}) \mathcal{O}_\eta(0) \rangle\,,
&
 \langle C_{\rm 2pt}^{\eta'}(t,\V{p}) \rangle
  &= \langle \mathcal{O}_{\eta'}(t;\V{p}) \mathcal{O}_{\eta'}(0) \rangle\, .
\label{eq:naive-twopoint}
\end{align}
The energy of the state $\alpha$ at a momentum
$\V{p}$, $E_{\alpha}(\V{p})$, can then be obtained
by fitting these two-point functions
at sufficiently large times $t$ to the functional form
\begin{equation}
\langle C_{\rm 2pt}^{\alpha}(t,\V{p}) \rangle
 =  A_{\alpha}(\V{p}) \left( \exp[-t E_{\alpha}(\V{p})] + \exp[-(T-t)E_{\alpha}(\V{p})]\right)\,,
 \label{eq:2ptfit}
\end{equation}
where $T$ is the temporal lattice extent,
and $A_{\alpha}(\V{p})$ is a (momentum-dependent) amplitude.

\begin{figure}
 \noindent 
\includegraphics{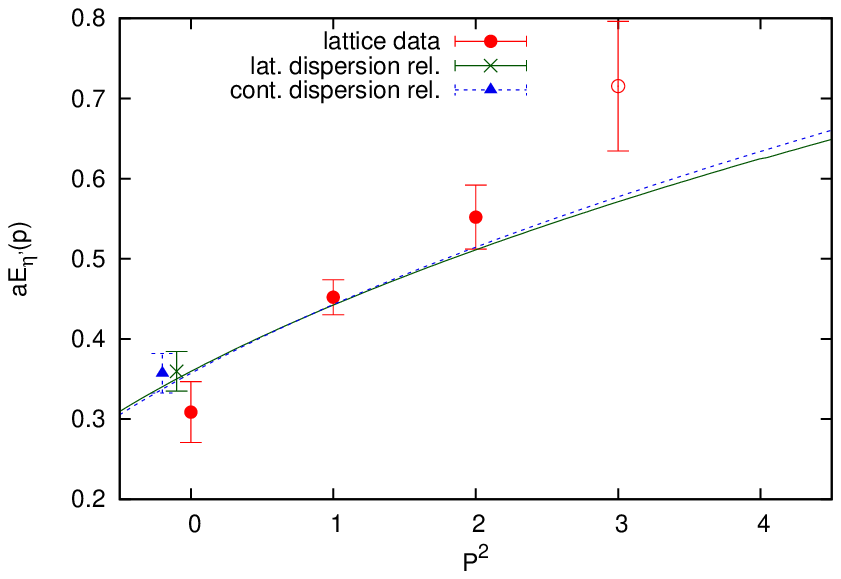}
\hfil
\includegraphics{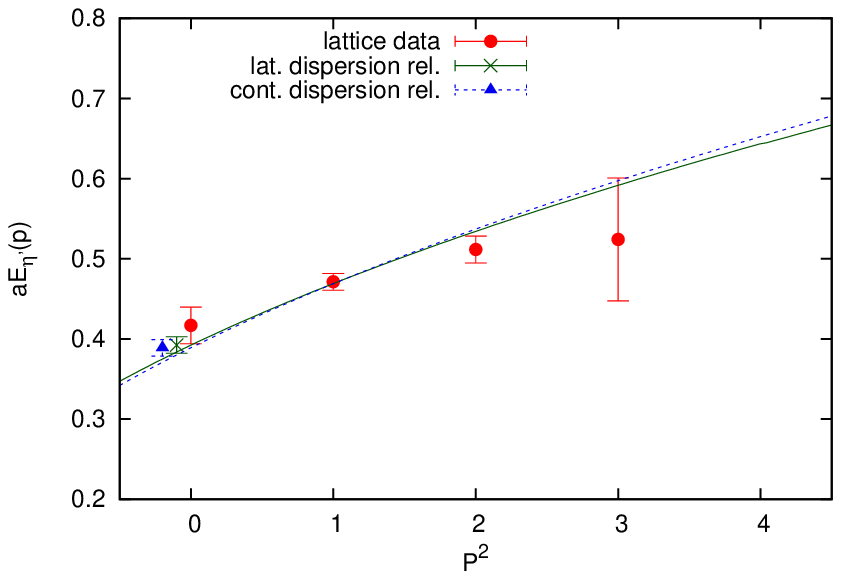}

 \noindent
\includegraphics{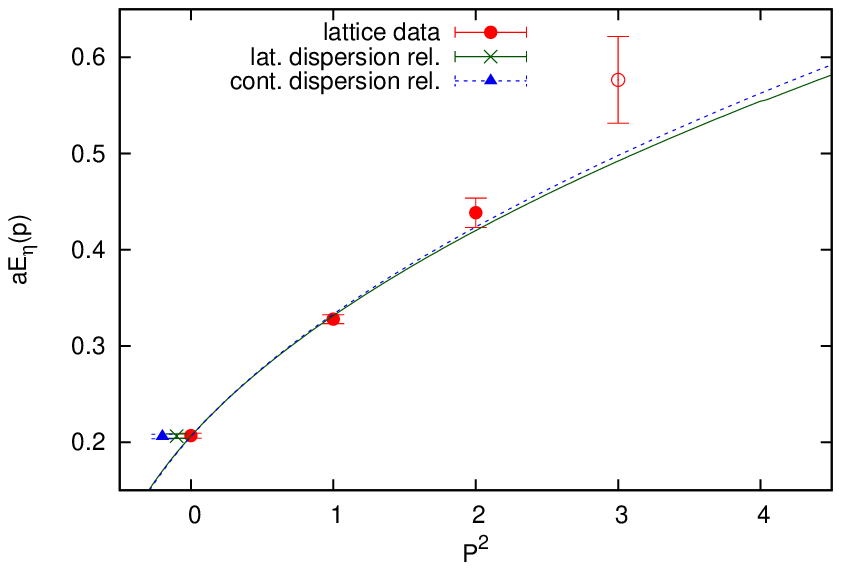}
\hfil
\includegraphics{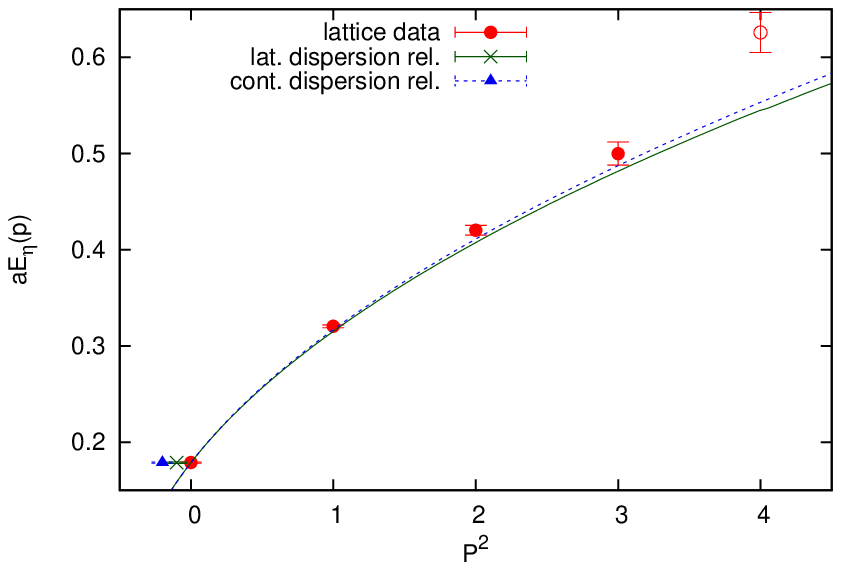}

 \noindent
\includegraphics{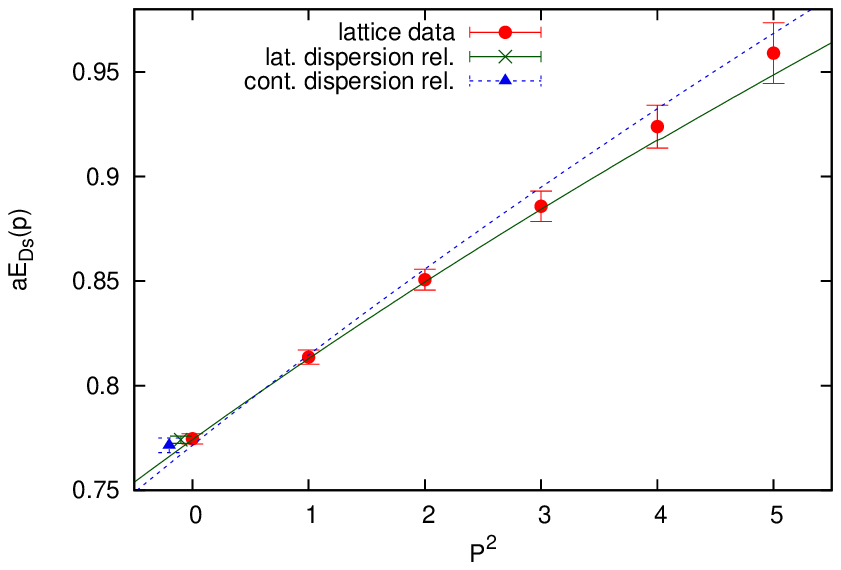}
\hfil
\includegraphics{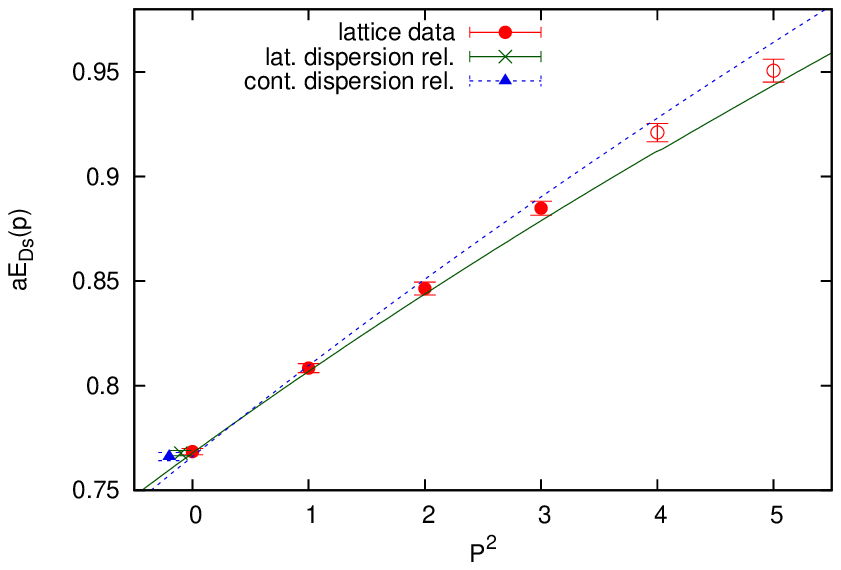}

 \caption{Energies $E(\V{p})$
along with fitted lattice and continuum dispersion relations
of the $\eta'$ (top),
 $\eta$ (middle) and $D_s$ (bottom) mesons
 at $M_\pi\approx 370$ MeV (Set A, left) and at the
SU(3) flavour symmetric 
point $M_\pi\approx 470$ MeV (Set S, right).
The momentum $\V{P}=\V{p} \times L/(2\pi)$ is in lattice units.
Open circles were excluded from the fits. 
$E(\V{p}=\V{0})$ for $\eta$ (Set A) and $\eta'$ (Set A, S) were obtained
by using the improved method explained in Sec.~\ref{sec:finite_volume}.
For clarity, the different mass determinations are replotted at
$\V{P}^2=0$ with a slight horizontal shift.
}
\label{fig:dispersion}
\end{figure}

At zero momentum the situation is more involved
and we deviate from the above procedure, see
Section~\ref{sec:finite_volume}. For the $\eta'$
mass on Set A and Set S and the $\eta$ mass on Set A,
the statistical error
of $M_{\alpha}=E_{\alpha}(\V{0})$ could be
further reduced by fitting zero- and non-zero-momentum data to
the lattice dispersion relation
\begin{equation}
 2\cosh(aE(\V{p})) = 2\cosh(a M) + \sum_{i=1}^3 4\sin^2\frac{a p_i}{2}\,,
\label{eq:lattice_dispersion}
\end{equation}
where the mass, $aM$, is a free parameter.
This is illustrated in
Fig.~\ref{fig:dispersion}, where the energies we obtained directly at
zero and at finite momenta, and the fitted dispersion relations
and their results are shown.
The masses are listed in Table~\ref{tab:mass},
and the energies at zero and finite momenta in Tables~\ref{tab:2pt_setA}
and \ref{tab:2pt_setS} of Appendix~\ref{app:data}.
The $\eta$ meson at the SU(3) flavour symmetric point (Set S) is identical to
the pion (and the kaon) and the precision of its mass did not benefit
from including non-zero momentum data. In this case we display
the $\V{p}=\V{0}$ result in the table.
No significant differences were found for the $\eta$ and
$\eta'$ masses if the continuum dispersion relation
$E^2=M^2+\V{p}^2$ was used
instead, as seen in the figure. 
For comparison, we also show the $D_s$ data in the figure.
For this meson, the lattice dispersion relation is clearly
preferred by the data: the $\chi^2/{\rm d.o.f}$
from the correlated fit were poor (4.7 for Set S and 2.2 for Set A)
and did not reproduce the data.  In Fig.~\ref{fig:dispersion},
uncorrelated fits are shown in these cases.

\begin{figure}
 \noindent
 \includegraphics[width=0.48\linewidth]{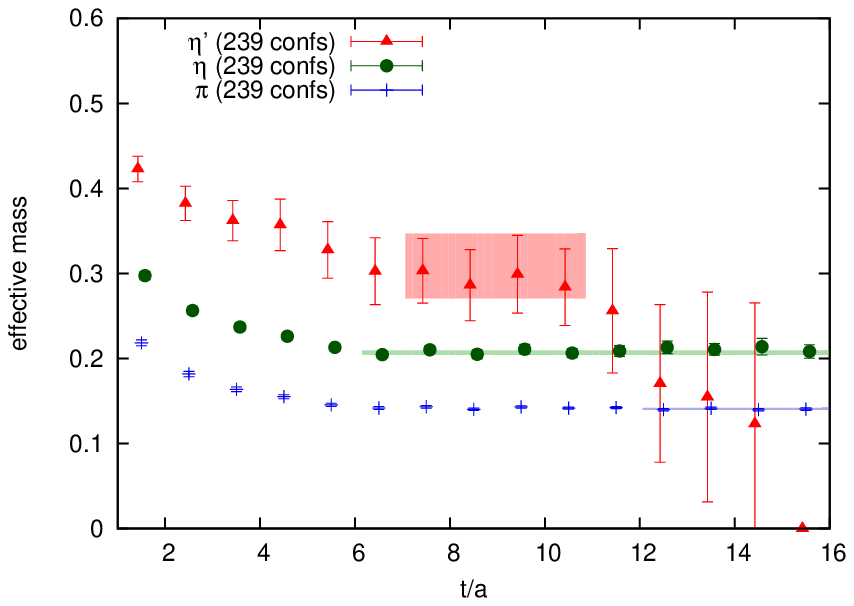}
 \hfil
 \includegraphics[width=0.48\linewidth]{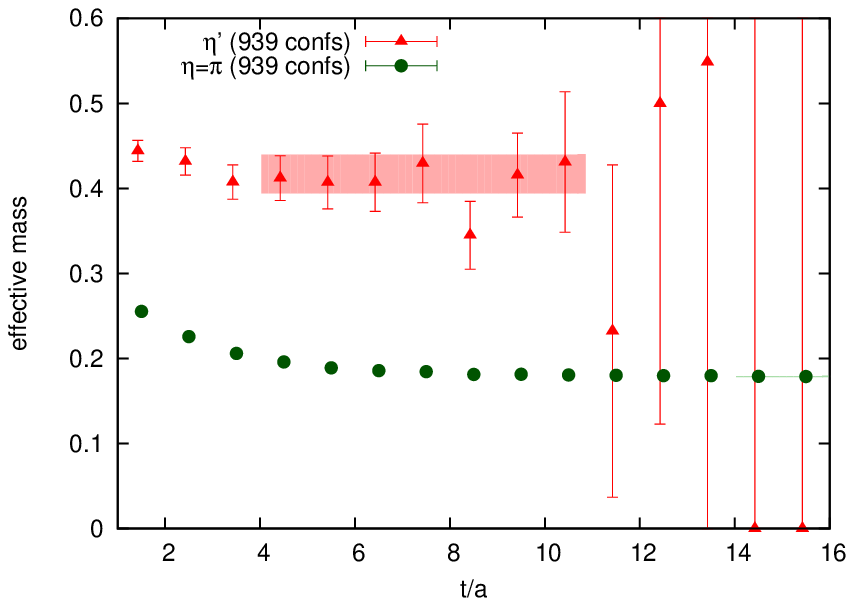}

\caption{Effective masses of the $\pi$, $\eta$
and  $\eta'$  for (left) $M_\pi \approx 370\, {\rm MeV}$~(Set A) and
(right) $M_\pi \approx 470\, {\rm MeV}$~(Set S).
Note that for Set S $\pi=\eta=\eta_8$ and $\eta'=\eta_1$.  
}
\label{fig:effmass}
\end{figure}

In Fig.~\ref{fig:effmass}, we plot the effective masses of the $\eta$,
the $\eta'$ and the $\pi$. The fitted $\eta$ and $\eta'$ masses,
with the exception of the $\eta=\eta_8=\pi$ at the SU(3) symmetric point, were
obtained using the improved method detailed in the next subsection.
The masses are
$M_\eta = 470.5 (1.8)\, {\rm MeV}$ and $M_{\eta'} = 1032(27)\,{\rm MeV}$
for Set S ($M_{\pi} \approx 470\,{\rm MeV} $),
$M_\eta = 542.8 (6.2)\, {\rm MeV}$ and $M_{\eta'} = 946(65)\,{\rm MeV}$
for Set A ($M_{\pi} \approx 370\,{\rm MeV} $),
where the errors are statistical only.
These values are consistent
with the finite momentum data shown in
Fig.~\ref{fig:dispersion}.

\begin{figure}
 \noindent
 \includegraphics{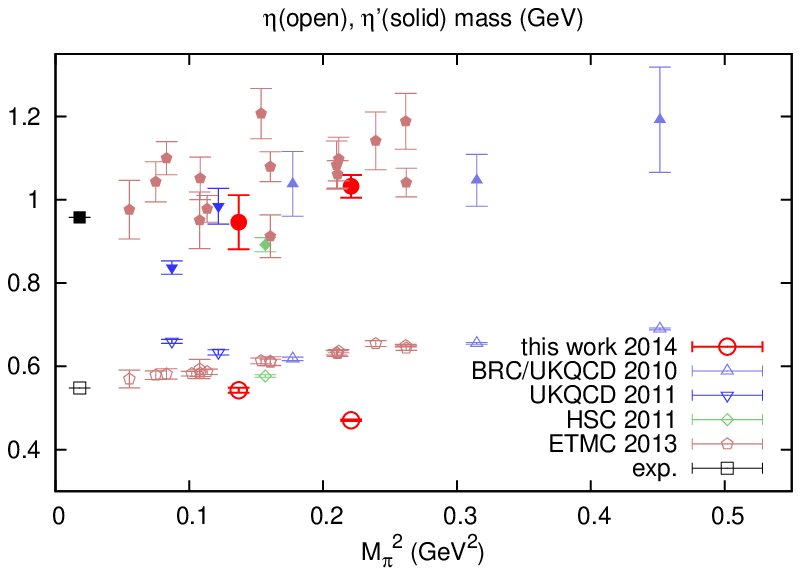}

\caption{Summary plot of recent lattice determinations of the 
$\eta$ (open symbols) and $\eta'$ (solid symbols) masses.
Results are shown from the
  RBC/UKQCD~(2010, \protect\cite{Christ:2010dd}),
  UKQCD~(2011, \protect\cite{Gregory:2011sg}),
  HSC~(2011, \protect\cite{Dudek:2011tt}),
 and
  ETMC~(2013, \protect\cite{Michael:2013gka})
 collaborations.
 The experimental values are taken from the Particle Data Group~\protect\cite{Agashe:2014kda}.
}
\label{fig:mass_summary}
\end{figure}

In Fig.~\ref{fig:mass_summary} we compare our $\eta$ and $\eta'$
masses to results obtained by other lattice collaborations
\cite{Christ:2010dd,Gregory:2011sg,Dudek:2011tt,Michael:2013gka} and
the respective experimental values \cite{Agashe:2014kda}.
In some of these studies the extrapolation to the physical point was
performed, however, for consistency we do not show extrapolated
values. Note that, since the flavour singlet quark mass
average is kept fixed in our simulations, the mass of the $\eta$
approaches the physical point from
below. Our results seem to approach the experimental values
and the $\eta'$ masses are consistent with other lattice determinations 
that were obtained
keeping the strange quark mass approximately constant.

\subsection{Finite volume effects on the $\eta$ and $\eta'$ masses}
\label{sec:finite_volume}

Analysing Set S we found that the $\eta'(=\eta_1)$ two-point function
at large times $t$ does not decay to zero (cf. Eq.~(\ref{eq:2ptfit}))
but instead saturates at a small non-zero value.
This phenomenon can be explained as a
finite volume effect, coupled to unrealistic
fluctuations of the topological charge, due to an insufficient
sampling of the topological sectors within our limited statistics.
We will see that this can be cured by defining an improved observable
which also reduces the variance of disconnected pseudoscalar two-point
functions in the case of a correctly sampled topological charge.

The disconnected contributions can be obtained by correlating pairs of
momentum-projected ``1-point loops''.
The sum over such a one-point loop is proportional
to the fermionic definition of the topological charge $Q_f$:
\begin{equation}
 \sum_{t} C_{\rm 1pt} (t,\V{p}=\V{0}) 
  =\sum_{t}\sum_{\V{x}} C_{\rm 1pt}(t,\V{x})= \alpha Q_f\approx\alpha Q\,,
 \label{eq:rho_F}
\end{equation}
where the (dimensionless) proportionality constant $\alpha$
will depend on the quark
mass, the smearing function and the normalization of the interpolator
and we assume the fermionic and gluonic definitions of the
topological charge to agree $Q\approx Q_f$. The above relation
suggests an approximate proportionality between the topological
charge density and the fermionic one-point loop:
$\rho(t,\V{x}) \simeq C_{\rm 1pt}(t,\V{x})/(\alpha a^4)$.

If the topological charge is fixed to $Q$,
point-point correlators of the topological charge density
$\rho(x)$ will remain
finite for large separations $|x|$ \cite{Aoki:2007ka}:
\begin{equation}
 \langle \rho(x) \rho(0) \rangle_Q 
 \to \frac{1}{V_4}\left( \frac{Q^2}{V_4} - \chi_t - \frac{c_4}{2\chi_t V_4}\right)
  +\cdots\, ,
 \label{eq:psuedoscalar-density}
\end{equation}
where $V_4$ is the physical four-volume, $\chi_t$ is the topological 
susceptibility, and the dimensionful kurtosis
$c_4$ parameterizes the leading deviations
from Gaussian fluctuations of $Q$.
Projecting the above expression
onto a fixed spatial momentum,
the constant term only affects the $\V{p}= \V{0}$ case.

Using $\rho(x) \simeq C_{\rm 1pt}(x)/(\alpha a^4)$,
we obtain the following estimate of the $\eta'$ two-point function,
which is the singlet two-point function (see Eq.~(\ref{eq:eta1-eta1})) in the 
SU(3) flavour symmetric case:
\begin{align}
C^{\eta'}_{\rm 2pt} (t,\V{p}=\V{0})
&=
  C_{\rm conn.}(t,\V{p}=\V{0})
  -3 \frac{a^4}{V_4} \sum_{t_0/a=0}^{T/a-1} 
      C_{\rm 1pt}(t+t_0,\V{p}=\V{0}) C_{\rm 1pt}(t_0,\V{p}=\V{0})
  \nonumber\\
&=
  C_{\rm conn.}(t,\V{p}=\V{0})
  -3 \frac{\alpha^2a^{12}}{V_4} \sum_{t_0/a=0}^{T/a-1} \sum_{\V{x},\V{x}_0}
      \rho(t+t_0,\V{x}) \rho(t_0,\V{x}_0)\,.
\end{align}
Here $C_{\rm conn.}$ is the quark-line connected part of the two-point function
and $T$ is the temporal extent of
the lattice.  By using Eq.~(\ref{eq:psuedoscalar-density})
and the observation that $c_4$ is negligible for our ensembles, 
we obtain
\begin{align}
\langle C^{\eta'}_{\rm 2pt} (t,\V{p}=\V{0}) \rangle_Q
\to
\frac{3\alpha^2a^5}{T} \left(\chi_t -\frac{Q^2}{V_4} \right)
\label{eq:etap-fixedQ}
\end{align}
for large $t$, resulting in
the prediction for the finite volume effect at $|Q|=0$:
\begin{equation}
\langle C^{\eta'}_{\rm 2pt} (t,\V{p}=\V{0}) \rangle_{Q=0}
     \to \frac{3\alpha^2a^5 \chi_t}{T}
\qquad ( m_l = m_s)\,.
\label{eq:etap_2pt_Q0}
\end{equation}
For the non-SU(3) flavour symmetric case, in principle,
both the singlet and octet parts of the
$\eta'$ two-point function should contribute to the constant.
However, using only the singlet part gives 
a good approximation because $\sin\theta'$ 
in Eq.~(\ref{eq:physical_interpolator}) is small.
The singlet-to-singlet contribution to the $\eta'$
two-point function is
\begin{align}
 \langle C_{\rm 2pt}^{\eta'}(t,\V{p}) \rangle
 &= \cos^2\theta' 
 \langle \mathcal{O}_1(t,\V{p}) \mathcal{O}^\dagger_1(0)\rangle
 + \cdots\,,
\end{align}
and we obtain
\begin{equation}
\langle C^{\eta'}_{\rm 2pt} (t,\V{p}=\V{0}) \rangle_{Q=0}
  \to \cos^2\theta'\frac{3\alpha^2a^5 \chi_t}{T}
\qquad ( m_l\neq m_s)\,,
\label{eq:etap_2pt_Q0_setA}
\end{equation}
where we used a flavour-averaged proportionality constant\footnote{
For each flavour $a=l,s$, we have 
$\sum_t C^a_{\rm 1pt}(t,\V{p}=\V{0}) = \alpha_a Q$,
where the proportionality constant depends on the flavour through the quark mass.
$\alpha$ in Eq.~(\ref{eq:alpha_flavour_averaged}) can be written as
$\alpha= (2 \alpha_l +\alpha_s)/3$.
}
\begin{equation}
 \sum_{t} \frac{1}{3}\left[
	   2C^l_{\rm 1pt} (t,\V{p}=\V{0})
	   +C^s_{\rm 1pt} (t,\V{p}=\V{0})
         \right]
   = \alpha Q\,.
\label{eq:alpha_flavour_averaged}
\end{equation}

\begin{figure}
\noindent
 \includegraphics{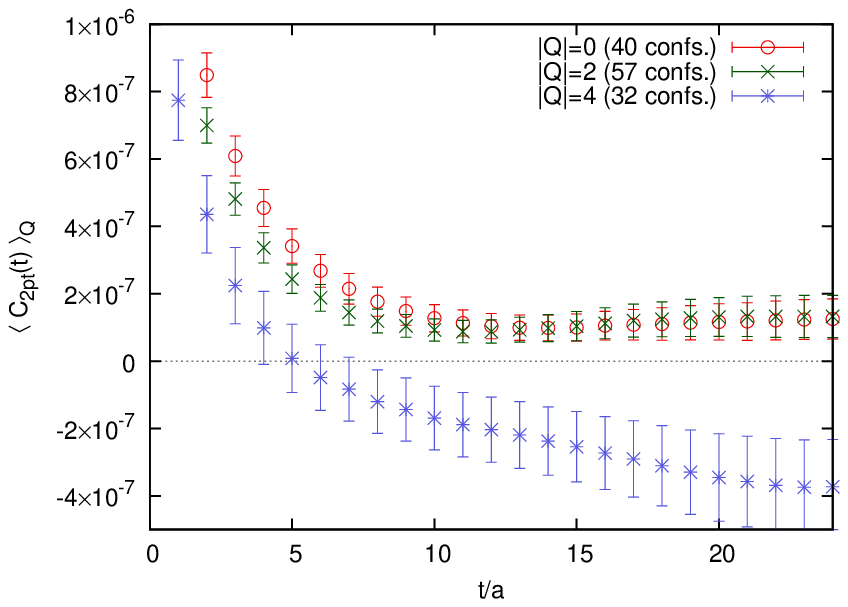}
\hfil 
 \includegraphics{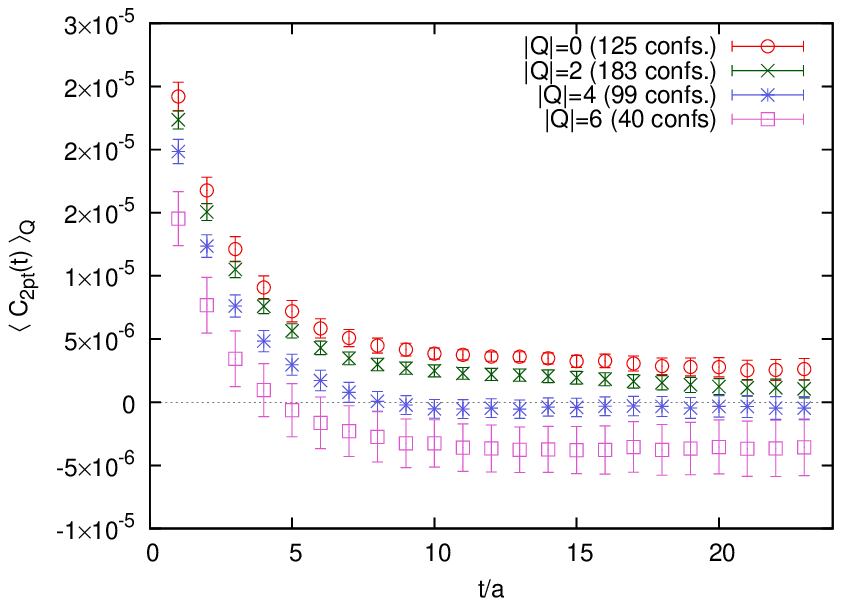}

 \caption{The naive zero-momentum $\eta'$ two-point function
 for each topological sector, for Set A (left panel) and Set S (right panel).}
 \label{fig:qdep_2pt}
\end{figure}

\begin{figure}
\noindent
 \includegraphics{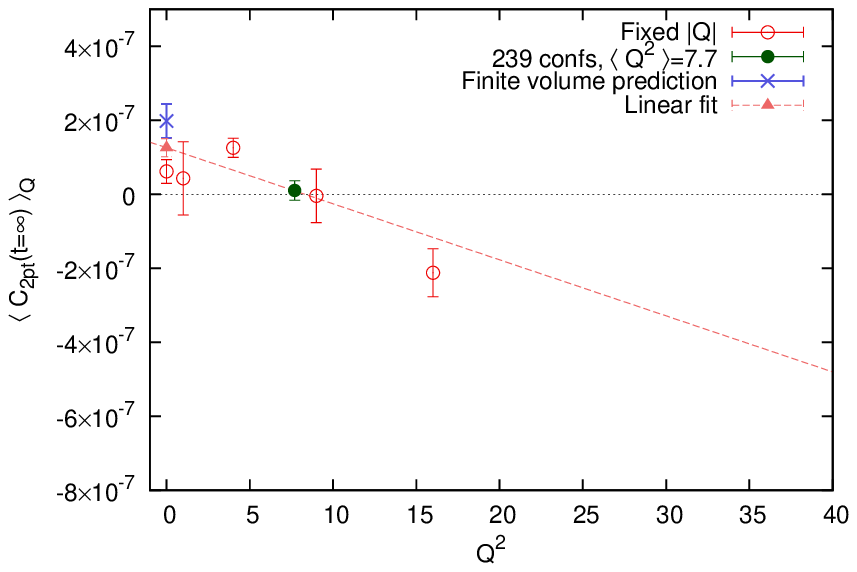}
\hfil
 \includegraphics{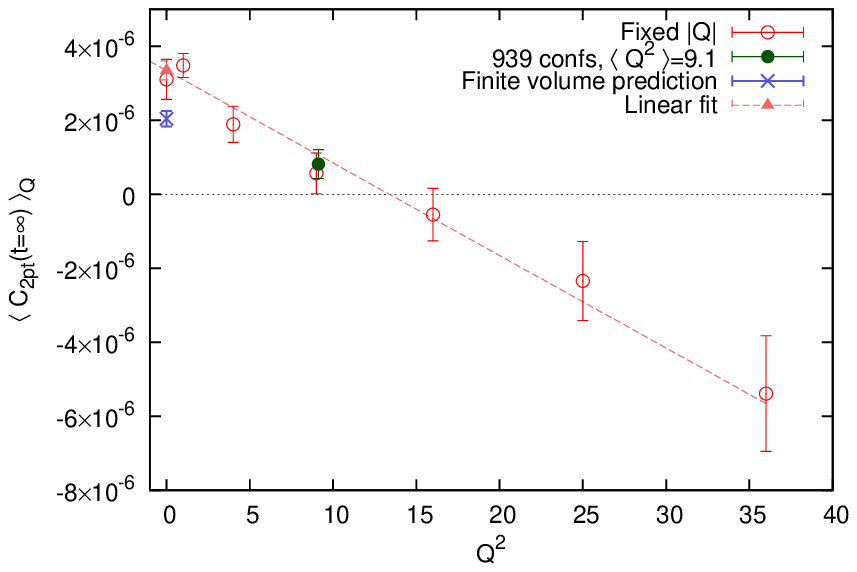}
 \caption{The constant part of the naive zero-momentum
$\eta'$ two-point function
for each topological sector, for Set A (left panel) and Set S (right panel).
The green solid circles were obtained
using all configurations. We found
$\langle Q^2 \rangle \approx 7.7$ for Set A and $\langle Q^2\rangle\approx 9.1$ for Set S. 
The $Q=0$ finite volume predictions (blue crosses,
Eqs.~(\protect\ref{eq:etap_2pt_Q0_setA})
and (\protect\ref{eq:etap_2pt_Q0})) were calculated using
$\chi_t=\langle Q^2\rangle/V_4$ and slopes $\alpha$,
determined via Eq.~(\protect\ref{eq:alpha_flavour_averaged}).
The dashed pink lines are linear fits to the fixed $|Q|$ data.
}
 \label{fig:qdep_const}
\end{figure}

To check Eq.~(\ref{eq:etap-fixedQ}), we measured $Q$ using an improved field
strength tensor \cite{BilsonThompson:2002jk}
on smeared gauge fields with 
90 iterations of Stout \cite{Morningstar:2003gk} smearing.
The measured values clustered around integer values as expected.
For each integer $n\geq 0$, using configurations with 
$n-0.5 \leq |Q|< n+0.5$ only (we denote them as $|Q|=n$ configurations),
we calculated the two-point function of the $\eta'$ at zero momentum.
The values of the two-point functions in the large time limit 
exhibit a clear dependence on
$|Q|$, see Fig.~\ref{fig:qdep_2pt}.  Moreover, the constants obtained by 
fitting within such subsets are consistent with
the linear dependence on $Q^2$
suggested by Eq.~(\ref{eq:etap-fixedQ}),
see Fig.~\ref{fig:qdep_const}.
See Ref.~\cite{Bali:2001gk} for an earlier observation of the $|Q|$ dependence
of the $\eta'$ effective mass, Ref.~\cite{Brower:2003yx} for the
general argument and Ref.~\cite{Kaneko:2009za} for a fixed topology approach.

In Fig.~\ref{fig:qdep_const}, we also plot the $Q=0$ predictions.
These were obtained from Eq.~(\ref{eq:etap_2pt_Q0}) (Set S) and
Eq.~(\ref{eq:etap_2pt_Q0_setA}) (Set A). The topological
susceptibilities $\chi_t=\langle Q^2\rangle/V_4$ were computed using
the gluonic definition of the topological charge and the parameters
$\alpha$ were obtained by fitting the one-point loops as a function of
the gluonic topological charge $Q$ as in 
Eq.~(\ref{eq:alpha_flavour_averaged}).  For Set A we find consistency between
the linear extrapolation and this $Q=0$ prediction. On Set S, however,
the prediction is significantly smaller than the extrapolated value,
and also smaller than the measured values.  
At the same time the $\eta'$ two-point function,
averaged over all configurations, approaches a non-zero (positive)
value. The linear fit of the constant part versus $Q^2$ crosses zero
at a value $\langle Q^2\rangle\approx 13$. Replacing the measured
value $\chi_t=\langle Q^2\rangle/V_4=9.1(0.4)/V_4$ within
Eq.~(\ref{eq:etap_2pt_Q0}) by $13/V_4$, the prediction would be
compatible with the fixed topology measurements. These observations
are coherent with our above arguments and strongly suggest $\langle
Q^2\rangle$ on Set S to be underestimated, due to an insufficient
sampling.

While the distribution of $Q$ on Set S is too narrow, we find
$\langle Q\rangle=0$ within errors on both ensembles.
Therefore, replacing $C_{\mathrm{1pt}}\mapsto
C_{\mathrm{1pt}} - \langle C_{\mathrm{1pt}}\rangle$ within
the above two-point functions will not affect any expectation value or
correct for the sampling of topological sectors in Set S.
Nevertheless, we checked whether this procedure reduced the statistical noise
but we did not find any improvement.

One way of addressing the problem of a non-vanishing expectation value
of the two-point function at large Euclidean times is simply to fit
the correlation function to a constant plus an exponential decay (which
we denote as ``naive fit with a constant'').  We adopted, however, a
different strategy that we found to reduce the gauge noise: 
this is motivated by the results in Fig.~\ref{fig:qdep_const}, which suggest that 
the two-point functions are shifted by different values in different
topological sectors according to Eq.~(\ref{eq:etap-fixedQ}). Therefore, normalizing the result 
to the $Q=0$ sector may reduce the gauge fluctuations.
We first
add a term that cancels the $Q^2$ dependence of
Eq.~(\ref{eq:psuedoscalar-density}), and then fit the result to a
constant plus an exponential decay (denoted as the ``improved
method'').

The details of the improved method are as follows.
Noting that the $Q^2$ term in Eq.~(\ref{eq:etap-fixedQ}) comes from
the disconnected part of the two-point function,
we replace this contribution to the two-point function
$D_{ab}(t)$ (see Eq.~(\ref{eq:disconn2pt}) of Appendix~\ref{app:2pt}) by
\begin{equation}
 D_{ab} (t) 
 = \frac{a^4}{V_4} \sum_{t_0/a=0}^{T/a-1}
   C_{\rm 1pt}^{a}(t+t_0) C_{\rm 1pt}^{b}(t_0)
 \mapsto
\tilde{D}_{ab}(t)
 \equiv
 D_{ab}(t)
 -\frac{a^5}{V_4T} \sum_{t_1/a, t_2/a=0}^{T/a-1}
   C_{\rm 1pt}^a(t_1) C_{\rm 1pt}^b(t_2)\,,
 \label{eq:disonn_replacement}
\end{equation}
where $\V{p}=\V{0}$ is understood. We perform this
subtraction on a configuration by
configuration basis, shifting the correlator
on different configurations by different values.
This results in a ``wrong'' expectation
value of $D$ (and thus of $C_{\rm 2pt}^{\eta'}$) but 
the subtraction does not affect its $t$-dependence.
The resulting two-point function should approximately
reproduce the behaviour Eqs.~(\ref{eq:etap_2pt_Q0}) and
(\ref{eq:etap_2pt_Q0_setA}) of the $Q=0$ sector.
Note that the cancellation cannot be perfect since,
instead of subtracting $\left\langle\sum_tC_{\mathrm{1pt}}(t)\right\rangle_Q^2$
within each fixed topology sector, in Eq.~(\ref{eq:disonn_replacement})
we subtract
$\left\langle\left[\sum_tC_{\mathrm{1pt}}(t)\right]^2\right\rangle$,
thereby neglecting
fluctuations of $\sum_t C_{\rm 1pt}(t)$ about $\alpha Q$.

We remark that even on ensembles with the correct distribution of
the topological charge we recommend to subtract this constant term from $D$,
(approximately) normalizing this to the $Q=0$ behaviour,
Eqs.~(\ref{eq:etap_2pt_Q0}) and
(\ref{eq:etap_2pt_Q0_setA}), since this construction,
as we will see below, significantly improves the
signal over noise ratio.

Replacing $D_{ab}(t)$ 
with $\tilde{D}_{ab}(t)$ in $C_{\rm 2pt}^{ij}(t)$, $i,j=1,8$, as advertised
above, we obtain modified two-point functions
$\tilde{C}_{\rm 2pt}^{ij}(t)$, see Eqs.~(\ref{eq:eta8-eta8})--(\ref{eq:eta1-eta8}):
\begin{align}
 \tilde{C}_{\rm 2pt}^{88}
 &= \frac{1}{3}(
     C_{ll} +C_{ss}
      - 2\tilde{D}_{ll}- 2\tilde{D}_{ss} + 2\tilde{D}_{ls} + 2\tilde{D}_{sl}
  )\,,
 \label{eq:improved_88}
\\
 \tilde{C}_{\rm 2pt}^{11}
 &= \frac{1}{3}(
     2C_{ll} +C_{ss}
      - 4\tilde{D}_{ll}- \tilde{D}_{ss} - 2\tilde{D}_{ls} - 2\tilde{D}_{sl}
  )\,,
 \label{eq:improved_11}
\\
 \tilde{C}_{\rm 2pt}^{18}
 &= \left(\tilde{C}_{\rm 2pt}^{81}(t) \right)^*
  = \frac{\sqrt{2}}{3}(
     C_{ll} - C_{ss}
      - 2\tilde{D}_{ll} + \tilde{D}_{ss} + 2\tilde{D}_{ls} - \tilde{D}_{sl}
  )\,,
 \label{eq:improved_18}
\end{align}
where $C_{ab}$ is a connected two-point function with flavour $a,b=l,s$
and we have suppressed the $t$-dependence.
Each modified two-point function still approximately
reproduces the constant term
Eq.~(\ref{eq:etap_2pt_Q0}).
Solving the generalized eigenvalue problem, we obtain eigenvectors
$(\cos\tilde{\theta}, -\sin\tilde{\theta})^T$ 
and 
$(\sin \tilde{\theta}', \cos\tilde{\theta}')^T$.
It is convenient to write the two-point functions in matrix notation:
\begin{equation}
\begin{pmatrix}
 \langle \tilde{C}_{\rm 2pt}^{\eta}(t) \rangle & 0 \\
 0 &  \langle \tilde{C}_{\rm 2pt}^{\eta'}(t) \rangle
 \end{pmatrix} 
 = U(\tilde{\theta}, \tilde{\theta}')
\begin{pmatrix}
 \langle \tilde{C}_{\rm 2pt}^{88} (t) \rangle &
 \langle \tilde{C}_{\rm 2pt}^{81} (t) \rangle \\
 \langle \tilde{C}_{\rm 2pt}^{18} (t) \rangle &
 \langle \tilde{C}_{\rm 2pt}^{11} (t) \rangle
\end{pmatrix}
   U^T(\tilde{\theta}, \tilde{\theta}')\,,
 \label{eq:diagonalization}
\end{equation}
where
\begin{equation}
 U(\tilde{\theta},\tilde{\theta}')\equiv
\begin{pmatrix}
 \cos \tilde{\theta} & -\sin\tilde{\theta} \\
  \sin\tilde{\theta}' &\cos\tilde{\theta}'
\end{pmatrix}\,.
\end{equation}
The modified two-point functions of the physical interpolators
at large times behave as
\begin{align}
 \left\langle \tilde{C}_{\rm 2pt}^{\eta}(t) \right\rangle
 &= A_\eta\left( \exp[-E_\eta t] + \exp[-E_\eta (T-t)] \right)+ \beta_\eta\,,
 \label{eq:etap_exp_plus_const0}
 \\
 \left\langle \tilde{C}_{\rm 2pt}^{\eta'}(t) \right\rangle
 &= A_{\eta'} \left(\exp[-E_{\eta'} t] + \exp[-E_{\eta'} (T-t)]  \right)+ \beta_{\eta'}\,.
 \label{eq:etap_exp_plus_const}
\end{align} 
From this we can obtain the constants $\beta_{_\eta}$ and $\beta_{\eta'}$.
At the SU(3) symmetric point, 
where $\tilde{\theta}=\tilde{\theta}'=0$,
we obtain the mass of the $\eta'$
from Eq.~(\ref{eq:etap_exp_plus_const}) alone. In this case
$\eta=\eta_8$ does not contain disconnected contributions
and $\beta_{\eta}=0$.

At the non-flavour symmetric point, the removal of the constant part
is more involved.
By inverting Eq.~(\ref{eq:diagonalization}) we can obtain
the contributions to $\beta_\eta$ and $\beta_{\eta'}$
from the two-point
functions in the octet-singlet basis.
We define improved
two-point functions for $\V{p}=\V{0}$, subtracting these:
\begin{equation}
  \begin{pmatrix}
  \langle C_{\rm 2pt}^{88}(t) \rangle & \langle C_{\rm 2pt}^{81}(t) \rangle \\
  \langle C_{\rm 2pt}^{18}(t) \rangle & \langle C_{\rm 2pt}^{11}(t) \rangle
 \end{pmatrix}_{\rm improved}
 =
  \begin{pmatrix}
  \langle \tilde{C}_{\rm 2pt}^{88}(t) \rangle & \langle \tilde{C}_{\rm 2pt}^{81}(t) \rangle\\
  \langle \tilde{C}_{\rm 2pt}^{18}(t) \rangle & \langle \tilde{C}_{\rm 2pt}^{11}(t) \rangle
 \end{pmatrix}
 -
 U^{-1}(\tilde{\theta}, \tilde{\theta}')
 \begin{pmatrix}
   \beta_\eta & 0 \\
   0 & \beta_{\eta'}
 \end{pmatrix}
 \left(U^{-1}\right)^T(\tilde{\theta}, \tilde{\theta}')\,.
\label{eq:improvedmixing}
\end{equation}
Solving the generalized eigenvalue problem for
the improved two-point functions,
we then obtain the masses and improved $\theta$ and $\theta'$
angles, that we will use to construct
the physical interpolators
at $\V{p}=\V{0}$. 

\begin{figure}
\noindent\hfil
 \includegraphics[width=0.48\linewidth]{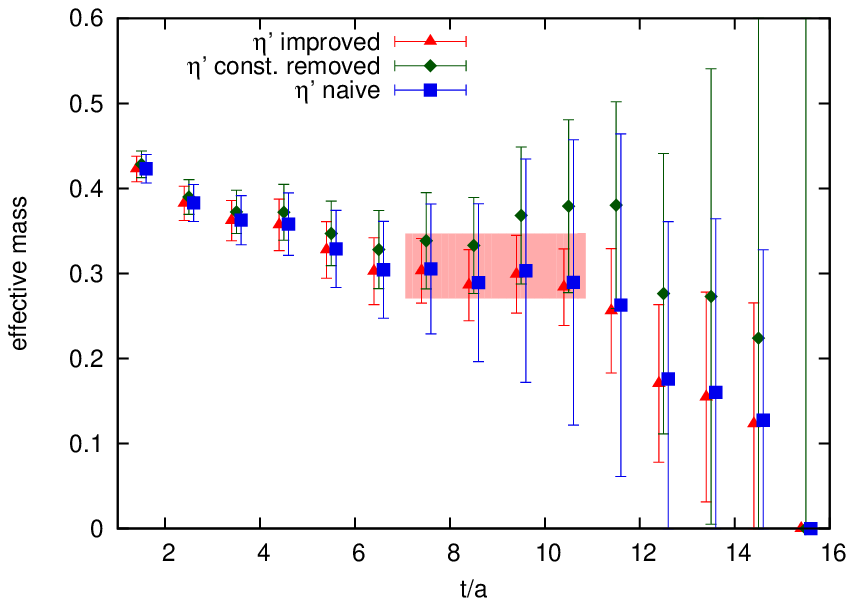}
 \hfil
 \includegraphics[width=0.48\linewidth]{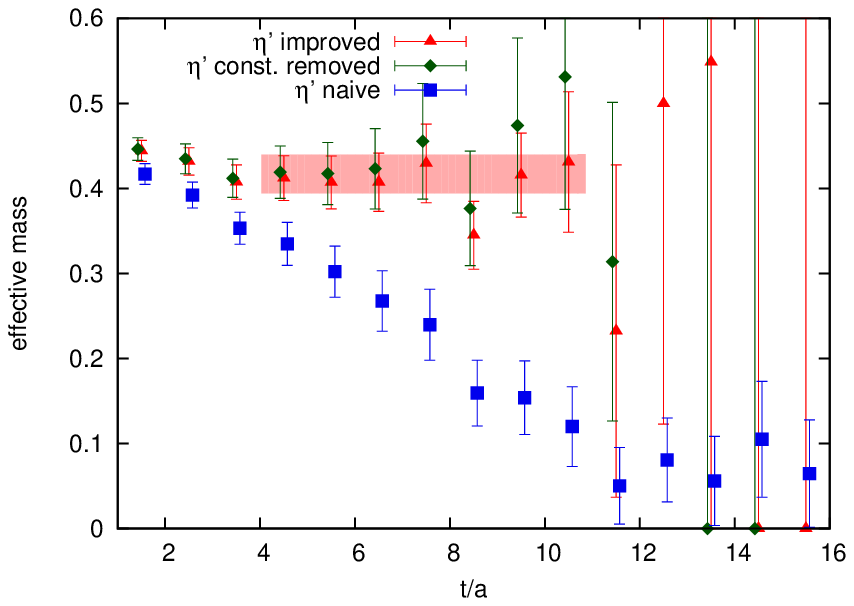}

 \caption{Effective mass of the $\eta'$, before and after the improvement,
   for Set A (left panel) and Set S (right panel).  Results are shown
   for three cases (a) using the naive $\eta^\prime$ two-point
   correlators, $C^{\eta^{(\prime)}}_{2pt}$, of
   Eq.~(\protect\ref{eq:naive-twopoint}) (blue squares), (b) using the naive
   correlators after removing the constant term (green diamonds) and
   (c) using the improved two-point functions, removing
the constant part Eq.~(\protect\ref{eq:etap_exp_plus_const})
for Set S, and solving the generalized
eigenvalue problem for Eq.~(\protect\ref{eq:improvedmixing}) 
   for Set A (red triangles).
}
\label{fig:effmass-before-after}.

\end{figure}

The effective masses of the $\eta'$ before and after the improvement are
plotted in Fig.~\ref{fig:effmass-before-after} for the two ensembles.
The results obtained from the naive fit with a constant are also
shown. For very large statistics there should be no difference between
the naive effective mass and the other two definitions, however,
as we have already discussed above, Set S showed a non-realistic
distribution of the topological charge.
The improved method gives the best signals and shows clear plateaus.

The method we presented here was motivated by the inadequate
sampling of the topological charge on one of our ensembles.
However, it is generally applicable to calculations of disconnected
contributions to light pseudoscalar two-point functions.
The improved two-point functions show reduced fluctuations,
at the price of a constant term that needs to be fitted.
In spite of this additional parameter, the extracted
masses are more precise than they are using the naive approach.

\begin{table}
\caption{Masses of the $\eta$ and the $\eta'$ mesons. The errors
are statistical only.
\label{tab:mass}}
\begin{ruledtabular}
  \begin{tabular}{cccc}
 Set  & $M_\eta$ [MeV] & $M_{\eta'}$ [MeV]\\
\hline
S & 470.5 (1.8) & 1032 (27) \\
A & 542.8 (6.2) & 946 (65)
 \end{tabular}
\end{ruledtabular}
\end{table}

\subsection{Mixing of the $\eta$ and $\eta'$ mesons in the octet-singlet basis}
\label{sec:mixing}

In addition to the mass, the mixing angles between the physical
$\eta^{\prime}$ and $\eta$ states and the octet-singlet basis are also
of phenomenological importance.
We restrict ourselves to Set A since at the
SU(3) flavour symmetric point (Set S) there is no such mixing and
$\eta'=\eta_1$, $\eta=\eta_8=\pi$. We define the two leading distribution
amplitudes
\begin{equation}
A_{j\eta^{(\prime)}}
 \equiv  \langle 0 | \mathcal{O}_j^{\rm local}|\eta^{(\prime)}\rangle\,,
\end{equation}
where $\mathcal{O}_j^{\rm local}$ is a local singlet ($j=1$) or octet ($j=8$)
interpolator projected onto zero momentum,
and use the following parameterizations
for which the renormalization factors of $\mathcal{O}_j^{\rm local}$
cancel \cite{Gregory:2011sg} (see also \cite{Feldmann:1999uf} and
references therein:\footnote{
Note that in Ref.~\cite{Feldmann:1999uf},
decay constants $f_{j\eta^{(\prime)}}$ are used 
instead of the $A_{j\eta^{(\prime)}}$, 
which are defined as 
$\langle 0 |A^{\mu}_j|\eta^{(\prime)}\rangle = ip^\mu f_{j\eta^{(\prime)}}$
with axial octet and singlet currents $A^\mu_j$.})
\begin{align}
 \frac{A_{8\eta'}}{A_{8\eta}}&= \tan \theta_8\, ,
&
 \frac{A_{1\eta}}{A_{1\eta'}}&= -\tan \theta_1\, ,
&
 \tan^2 \bar{\theta} = \tan\theta_8 \tan\theta_1\,.
\label{eq:thetas}
\end{align}
To obtain these amplitudes, 
we use the asymptotic behaviour at large times $t$ of 
smeared source to local~(point) sink two-point functions at zero
momentum\footnote{%
Note that we
use the improved method outlined in the previous subsection,
Eqs.~(\ref{eq:improved_88})--(\ref{eq:improved_18}), replacing the
disconnected contribution as in Eq.~(\ref{eq:disonn_replacement}),
this time also for smeared-point two-point functions.
Therefore, we have to allow for constant contributions that
we denote as $\beta_{j\eta^{(\prime)}}$.
}
\begin{align}
 \left\langle C_{\rm 2pt, sm\to pt}^{j\eta^{(\prime)}}(t) \right\rangle
=
 \langle 0 |\mathcal{O}_j^{\rm local}(t) 
\mathcal{O}_{\rm \eta^{(\prime)}}^\dagger(0) |0\rangle
&\to \frac{A_{j\eta^{(\prime)}}Z_{\eta^{(\prime)}}}{2M_{\eta^{(\prime)}}}
   \left(\exp[-M_{\eta^{(\prime)}}t] + \exp[-M_{\eta^{(\prime)}}(T-t)]\right)
 +\beta_{j\eta^{(\prime)}}\,,
 \label{eq:fit_ampl}
\end{align}
where $Z_{\eta^{(\prime)}} =\langle
\eta^{(\prime)}|\mathcal{O}_{\eta^{(\prime)}}^\dagger|0\rangle $ (see
Eq.~(\ref{eq:amplitude}) below) can be obtained from the
smeared-smeared two-point function.
The physical $\eta$ or $\eta'$ state is
created by $\mathcal{O}_{\eta^{(\prime)}}^\dagger$, for which we use
the improved $\theta$ or $\theta'$ parameters obtained from the
smeared-smeared
correlators in the previous subsection. 
Note that these angles depend on our choice of
smearing and --- unlike the mixing angles discussed below ---
are not properties of the physical states alone.
Using the mixing angles $\theta$ and $\theta'$ we build the improved
two point functions
\begin{align}
\left\langle \tilde{C}_{\rm 2pt, sm\to pt}^{j\eta}(t) \right\rangle 
 & = 
 \cos\theta  
 \left\langle \tilde{C}_{\rm 2pt, sm\to pt}^{j8}(t)\right\rangle 
 -\sin\theta 
 \left\langle \tilde{C}_{\rm 2pt, sm\to pt}^{j1}(t)\right\rangle \,,\\
\left\langle \tilde{C}_{\rm 2pt, sm\to pt}^{j\eta^{\prime}}(t) \right\rangle
 & = 
 \sin\theta^\prime
 \left\langle \tilde{C}_{\rm 2pt, sm\to pt}^{j8}(t)\right\rangle
 +\cos\theta^\prime 
 \left\langle\tilde{C}_{\rm 2pt, sm\to pt}^{j1}(t)\right\rangle\,.
\end{align}
Both sides of the above equations may contain
constant contributions, due to the replacement $C\mapsto \tilde{C}$
coming from the improved method.

We fit $\langle \tilde{C}_{\rm 2pt, sm\to pt}^{j\eta^{(\prime)}}(t) \rangle$,
fixing the mass $M_{\eta^{(\prime)}}$ to the value we determined
previously, leaving $A_{j\eta^{(\prime)}}$ and
$\beta_{j\eta^{(\prime)}}$ as free parameters. The resulting
angles $\theta_{1}$, $\theta_8$ and $\bar{\theta}$,
see Eq.~(\ref{eq:thetas}), are given
in Table~\ref{tab:mixing}.  The first error is statistical, while the
second one is an estimate of the systematics from the choice
of the fit range and was obtained varying this by $\pm 1$
timeslices.  Note that, since both $\tan \theta_8$ and $\tan \theta_1$
are negative, we also adopted a negative value for $\tan\bar{\theta}$.
$\theta_8$ was found to differ from $\theta_1$ (and hence from
$\bar\theta$): two angles are needed to connect the
physical states to the octet-singlet basis, indicating
the relevance of higher Fock states.
A phenomenological estimate used in Ref.~\cite{Feldmann:1999uf}
also gives two mixing angles, 
$\theta_8=-21.2(1.6)^\circ$ and $\theta_1=-9.2(1.7)^\circ$,
where the errors are solely experimental
and no systematic errors are included.
In the lattice study of Ref.~\cite{Christ:2010dd} a single mixing angle
$-14.1(2.8)^\circ$ was obtained, relative to the octet-singlet basis,
after extrapolating to the physical point. This is in the middle
between the phenomenological $\theta_1$ and $\theta_8$ values.
The ratio $\theta_8/\theta_1$ of Ref.~\cite{Feldmann:1999uf}
is consistent with our result, however,
both our angles come out a factor of two smaller than in that analysis.
This is not surprising since we start from the flavour-symmetric
point where  $\theta_8=\theta_1=0$, while Set A corresponds to
a quark mass ratio $m_s/m_l\approx 2.8$, still quite far away
from the physical point $m_s/m_l\approx 25$.
A monotonous extrapolation would indeed suggest larger values
of $|\theta_j|$ for physical $m_s/m_l$.

Another interesting combination are ratios of the
$A_{j\eta^{(\prime)}}$ amplitudes to a similar distribution amplitude
for the pion
\begin{equation}
 \frac{A_{j\eta^{(\prime)}}}{A_\pi} 
 \qquad
 \text{with}
 \qquad
A_\pi\equiv \langle 0 | \mathcal{O}_\pi^{\rm local} |\pi\rangle,
\end{equation}
where $\mathcal{O}_\pi^{\rm local}$ is the local pion
interpolator.
Note that the renormalization factors only cancel exactly
for the ratios $A_{8\eta^{(\prime)}}/A_{\pi}$  while in the singlet
case this is violated at
two-loop order in perturbation theory.
In Table~\ref{tab:distrib_amplitude}, we list the values (in the left column).
The octet component of the $\eta$ meson is
enhanced, relative to the flavour-symmetric case
while the singlet $\eta'$ distribution amplitude
is much smaller than that for the pion.
Note that the negative value of $A_{8\eta'}$ signals
an octet-admixture to $\eta'$ much bigger than the
singlet component of $\eta$, which is another manifestation
of the result $|\theta_8|>|\theta_1|$.

For completeness, we also determined the angles and ratios using
the~(unimproved) $\langle C_{\rm 2pt, sm\to pt}^{j\eta^{(\prime)}}(t) \rangle$ with
both $\beta_{j\eta^{(\prime)}}=0$ and $\beta_{j\eta^{(\prime)}}\neq0$
in the fit function. The results are included in
Tables~\ref{tab:mixing} and \ref{tab:distrib_amplitude} for comparison. 
The three determinations are broadly consistent for both quantities. 
We see no significant
reduction in the statistical errors between the unimproved/improved
$\beta_{j\eta^{(\prime)}}\neq0$ cases. This may be due to the
use of the same~(improved)
$\theta$ and $\theta'$ to construct the physical states or
that the assumption of small fluctuations of $\sum_t C_{\rm
  1pt}(t)$ around the topological charge (see the argument below
Eq.~(\ref{eq:disonn_replacement})) may be less valid for the local
one-point loop. The naive (unimproved $\beta_{j\eta^{(\prime)}}=0$)
errors are slightly smaller since the fit parameters
$\beta_{j\eta^{(\prime)}}$ are fixed. The discussion
of the previous subsection, however, suggests that due to
the coupling between the disconnected loop and
the slowly moving topological charge
it is safer to allow for such constants.

\begin{table}
\caption{The mixing angles $\theta_8$,
 $\theta_1$ and $\bar{\theta}$ in degrees for Set A.
 The improved and unimproved values were obtained using
$\tilde{C}_{\rm 2pt, sm\to pt}^{j\eta^{(\prime)}}(t)$  and
$C_{\rm 2pt, sm\to pt}^{j\eta^{(\prime)}}(t)$, respectively.
The first errors are statistical and the second quantify the uncertainty
from the choice of the fit range.
}
 \label{tab:mixing}
 \begin{ruledtabular}
  \begin{tabular}{cccc}
   & improved, $\beta_{j\eta^{(\prime)}}\neq0$ & unimproved, $\beta_{j\eta^{(\prime)}}\neq0$ & unimproved, $\beta_{j\eta^{(\prime)}}=0$ \\
 \hline
 $\theta_8$ & $-10.9(1.5)_{\rm stat.}(0.5)_{\rm fit}$ 
            & $-10.9(1.5)_{\rm stat.}(0.4)_{\rm fit}$ 
            & $-10.5(1.1)_{\rm stat.}(0.2)_{\rm fit}$ \\
 $\theta_1$ & $-5.5(1.5)_{\rm stat.}(1.2)_{\rm fit}$ 
            & $-5.5(1.5)_{\rm stat.}(1.2)_{\rm fit}$
	    & $-7.1(1.2)_{\rm stat.}(1.3)_{\rm fit}$  \\
 $\bar{\theta}$ & $-7.7(0.9)_{\rm stat.}(0.8)_{\rm fit}$
                & $-7.7(0.9)_{\rm stat.}(0.7)_{\rm fit}$ 
	        & $-8.6(0.9)_{\rm stat.}(0.9)_{\rm fit}$
  \end{tabular}
 \end{ruledtabular}
\end{table}

\begin{table}
\caption{The distribution amplitudes for Set A normalized with respect to
$A_\pi$, using the three different methods. The (unknown) renormalization
factor exactly cancels from the octet ratios and is partially
canceled in the singlet ratios.
}
 \label{tab:distrib_amplitude}
 \begin{ruledtabular}
  \begin{tabular}{cccc}
   & improved, $\beta_{j\eta^{(\prime)}}\neq0$ & unimproved, $\beta_{j\eta^{(\prime)}}\neq0$ & unimproved, $\beta_{j\eta^{(\prime)}}=0$ \\
 \hline
 $A_{8\eta}/A_\pi$ & $1.124(14)_{\rm stat.}(04)_{\rm fit}$  
                   & $1.124(14)_{\rm stat.}(04)_{\rm fit}$
	           & $1.120(14)_{\rm stat.}(03)_{\rm fit}$ \\
 $A_{1\eta}/A_\pi$ & $0.058(24)_{\rm stat.}(12)_{\rm fit}$
                   & $0.058(24)_{\rm stat.}(12)_{\rm fit}$
                   & $0.082(19)_{\rm stat.}(08)_{\rm fit}$ \\
 $A_{8\eta'}/A_\pi$ & $-0.216(31)_{\rm stat.}(11)_{\rm fit}$ 
                    & $-0.216(31)_{\rm stat.}(11)_{\rm fit}$ 
                    & $-0.207(22)_{\rm stat.}(39)_{\rm fit}$\\
 $A_{1\eta'}/A_\pi$ & $0.60(13)_{\rm stat.}(20)_{\rm fit}$
                    & $0.60(13)_{\rm stat.}(21)_{\rm fit}$
	            & $0.65(12)_{\rm stat.}(17)_{\rm fit}$
  \end{tabular}
 \end{ruledtabular}
\end{table}

\section{Determination of the semileptonic form factors}
\label{sec:form factor}

Having obtained the $\eta$ and $\eta'$ interpolators, we are now
in the position to calculate the $D_s\rightarrow\eta\ell\bar{\nu}_{\ell}$
and $D_s\rightarrow\eta'\ell\bar{\nu}_{\ell}$ semileptonic decay
form factors $f_0(q^2)$. We discuss the relevant matrix
elements and our methods to compute these, before we present
and discuss our results on the form factors.
\subsection{Matrix elements}
\label{sec:matrix-elements}

The matrix elements needed to study the decays
$D_s \to \eta^{(\prime)}\ell \bar{\nu}_{\ell}$ are obtained 
from the following three-point functions:
\begin{equation}
 \left\langle 
  C_{\rm 3pt}^{D_s\to \eta^{(\prime)}}(t,\V{p},\V{k}; t_{\rm sep}) 
 \right\rangle
 = \langle 0| \mathcal{O}_{\eta(')}(\V{k}, t_{\rm sep}) S(\V{0},t)
\mathcal{O}_{D_s}^\dagger(\V{p},0) | 0 \rangle\,,
 \label{eq:3pt}
\end{equation}
where we used smeared interpolators $\mathcal{O}_{D_s}$ and 
$\mathcal{O}_{\eta^{(\prime)}}$ for both the $D_s$ and the
$\eta^{(\prime)}$, respectively.
$S$ is the local scalar current in position
space. It can also be averaged over the spatial volume (multiplying by
the phases $e^{i\V{q}\cdot(\V{x}-\V{x}_0)}$), to increase
statistics.
We detail the computation methods both for the connected and the
disconnected contributions 
to the three-point functions in Appendix~\ref{app:3pt}.
Fig.~\ref{fig:conn_vs_disconn} shows the full three-point function
and the contributions from connected and disconnected fermion loop diagrams.
It is interesting to note that the magnitude of the disconnected
contributions is large, especially for the decay to $\eta'$.
Not surprisingly, the statistical error of the three-point
function mainly comes from the disconnected part.

\begin{figure}
\noindent
\includegraphics{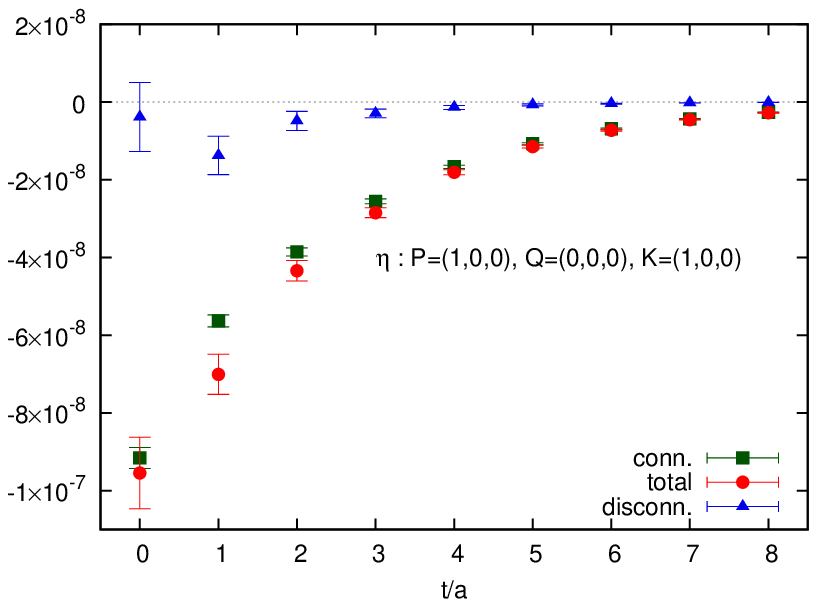}
\hfil
\includegraphics{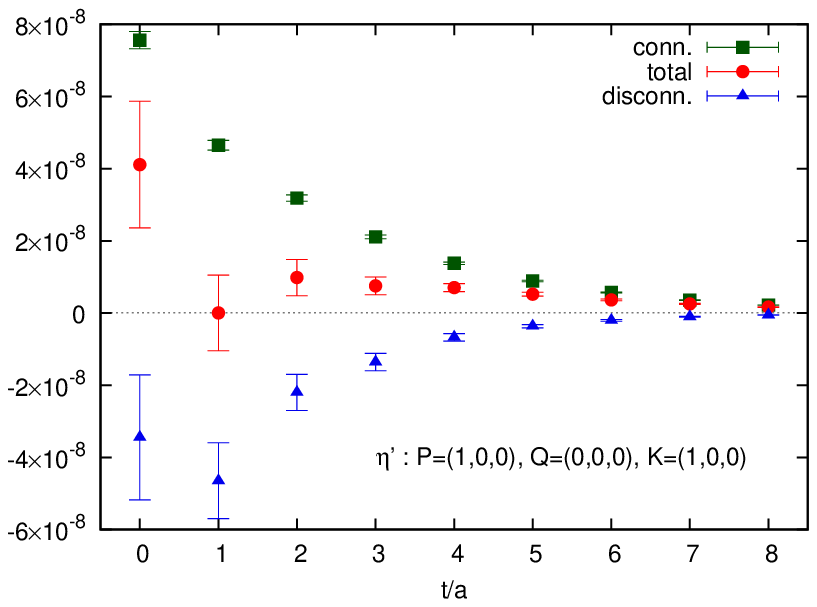}

\caption{Connected (green squares) and disconnected (blue triangles) 
contributions to the total three-point
 functions (red circles) for $D_s\to \eta$ (left panel) and
 $D_s \to \eta'$ (right panel) matrix elements for Set A with $t_{\rm sep}=8a$. 
The $D_s$ is located at $t=0$ and the $\eta^{(\prime)}$ is at $t=8a$.
 The momenta are in the lattice units: 
 $(\V{P}, \V{Q}, \V{K}) = (\V{p}, \V{q}, \V{k}) \times L/(2\pi)$.
}
\label{fig:conn_vs_disconn}
\end{figure}

The three-point functions
have the following spectral decomposition:
\begin{align}
 \lefteqn{
 \left\langle C_{\rm 3pt}^{D_s\to\eta^{(\prime)}}
  (t,\V{p},\V{k}; t_{\rm sep})  \right\rangle
}\qquad \nonumber\\
 &= \frac{Z_{\eta^{(\prime)}}}{2E_{\eta^{(\prime)}}} \frac{Z_{D_s}}{2E_{D_s}} 
    \langle \eta^{(\prime)}(\V{k}) |S(\V{0}) |D_s(\V{p})\rangle
    \exp\left[-E_{D_s}(T-t) - E_{\eta^{(\prime)}}(T-(t_{\rm sep} -t))\right]
   \nonumber \\
 &\quad 
    + 
    \frac{Z_{\eta^{(\prime)*}}}{2E_{\eta^{(\prime)*}}} \frac{Z_{D_s}}{2E_{D_s}} 
    \langle \eta^{(\prime)*}(\V{k}) |S(\V{0}) |D_s(\V{p})\rangle
    \exp\left[-E_{D_s}(T-t) - E_{\eta^{(\prime)*}}(T-(t_{\rm sep} -t))\right]
   \nonumber \\
 &\quad
    + 
    \frac{Z_{\eta^{(\prime)}}}{2E_{\eta^{(\prime)}}} \frac{Z_{D_s^*}}{2E_{D_s^*}} 
    \langle \eta^{(\prime)}(\V{k}) |S(\V{0}) |D_s^*(\V{p})\rangle
    \exp\left[-E_{D_s^*}(T-t) - E_{\eta^{(\prime)}}(T-(t_{\rm sep} -t))\right]
   \nonumber \\
 &\quad
    + 
    \frac{Z_{\eta^{(\prime)*}}}{2E_{\eta^{(\prime)*}}} \frac{Z_{D_s^*}}{2E_{D_s^*}} 
    \langle \eta^{(\prime)*}(\V{k}) |S(\V{0}) |D_s^*(\V{p})\rangle
    \exp\left[-E_{D_s^*}(T-t) - E_{\eta^{(\prime)*}}(T-(t_{\rm sep} -t))\right]
    \nonumber \\
 &\quad
    + \cdots\,,
\end{align}
where $*$ indicates the first excited states and 
we have neglected contributions from even higher excitations.
$Z_{X^{(*)}}$ is the amplitude of the state
with $X=D_s$, $\eta$ and $\eta'$.
For brevity we suppress the momentum dependence of $Z_{X^{(*)}}=Z_{X^{(*)}}(\V{p})$
and $E_{X^{(*)}}=E_{X^{(*)}}(\V{p})$.
The first term on the r.h.s contains the ground state to ground state
matrix element that we are interested in.

Note that the phase of the state $X$ is arbitrary and we choose it
such that we have a real positive amplitude
\begin{equation}
 Z_{X} = \langle X | \mathcal{O}_X^\dagger | 0\rangle > 0\,.
 \label{eq:amplitude}
\end{equation} 
This means that the matrix elements
$\langle D_s(\V{p})|S(\V{0})|\eta^{(\prime)} (\V{k}) \rangle$
can be negative\footnote{
Charge conjugation invariance guarantees the matrix element is real in 
coordinate space, and then parity invariance 
$\langle D_s(\V{p})|S(\V{0})|\eta^{(\prime)} (\V{k}) \rangle
 =\langle D_s(-\V{p})|S(\V{0})|\eta^{(\prime)} (-\V{k}) \rangle
$
gives a real three-point function in momentum space} and, indeed, we obtained negative values for the $\eta$.
Since the sign of the matrix element is not
physical, in the following we use its modulus.\footnote{Note, however,
that relative signs are relevant for studies of flavour mixing angles
in decays. This is similar to the connection of the
sign of the distribution amplitude ratio $A_{8\eta'}/A_\pi$
to the sign of the respective mixing angle $\theta_8$.}

In order to determine the ground state to ground state matrix element
reliably, it is important to take into account the excited state
contributions to the three-point function. One way to do this is to use a large
sink-source separation so that the excited state contributions are
small. However, this is not possible in the current case because
the statistical error grows rapidly (due to the disconnected
terms), even for relatively small time separations.  We need to employ an
alternative approach.

First we obtain
$E_X$ and $Z_X$ by fitting the 
two-point function
\begin{equation}
 \langle C_{X}^{\rm 2pt}(\V{p},t) \rangle
  = \frac{|Z_X|^2}{2E_X}\left( \exp[-E_X t] + \exp[-E_X(T-t)]\right)
  + \frac{|Z_{X^*}|^2}{2E_{X^*}}\left( \exp[-E_{X^*} t] + \exp[-E_{X^*}(T-t)]\right)
    +\cdots\,,  \label{twopoint-fit}
\end{equation}
using a functional form given by the first term, at sufficiently large $t$.
The energy gap, $\Delta E_X = E_{X^*} -E_X$, is  then determined by
fitting the combination
\begin{equation}
 \frac{\langle C^{\rm 2pt}_X(t,\V{p}) \rangle}{ \frac{|Z_{X}|^2}{2E_X}
  \left(\exp[-E_Xt] + \exp[-E_X(T-t)] \right)} -1
 \label{eq:def_DeltaE}
\end{equation}
to the form $a_X \exp(-\Delta E_X t)$, where not only
$\Delta E_X$ but also the amplitudes $a_X$ depend on the momentum $\V{p}$.
To extract the matrix element,
$\langle \eta^{(\prime)}(\V{k}) |S(\V{0}) |D_s(\V{p})\rangle$,
we compute the ratio
\begin{align}
R(t)
&=\frac{\langle C_{\rm 3pt}^{D_s \to \eta^{(\prime)}}
(t,\V{p},\V{k}; t_{\rm sep})\rangle}{
 \frac{Z_{D_s}}{2E_{D_s}}
 \left(
     \exp\left[-E_{D_s}t \right]
    +\exp\left[ -E_{D_s}(T-t) \right]
 \right)
   \frac{ Z_{\eta^{(\prime)}} }{ 2E_{\eta^{(\prime)}} }
     \left(
       \exp\left[ -E_{\eta^{(\prime)}}( t_{\rm sep} -t ) \right]
      +\exp\left[ -E_{\eta^{(\prime)}}( T-(t_{\rm sep} -t) )\right]
      \right) } 
\end{align}
and use the fit function
\begin{equation}
R(t)=
 c+ A_1 \exp\left[ -\Delta E_{D_s}t \right]
  + A_2 \exp\left[ -\Delta E_{\eta^{(\prime)}}(t_{\rm sep}-t) \right]\,,
\label{eq:const_plus_excited}
\end{equation}
where $c=\langle \eta^{(\prime)}(\V{k}) |S(\V{0}) |D_s(\V{p})\rangle$.
Whenever the two-point function had a small overlap with the excited state 
and we were unable to extract $\Delta E_{\eta^{(\prime)}}$,
we only employed the first two terms of Eq.~(\ref{eq:const_plus_excited}).

\begin{figure}
\noindent
 \includegraphics{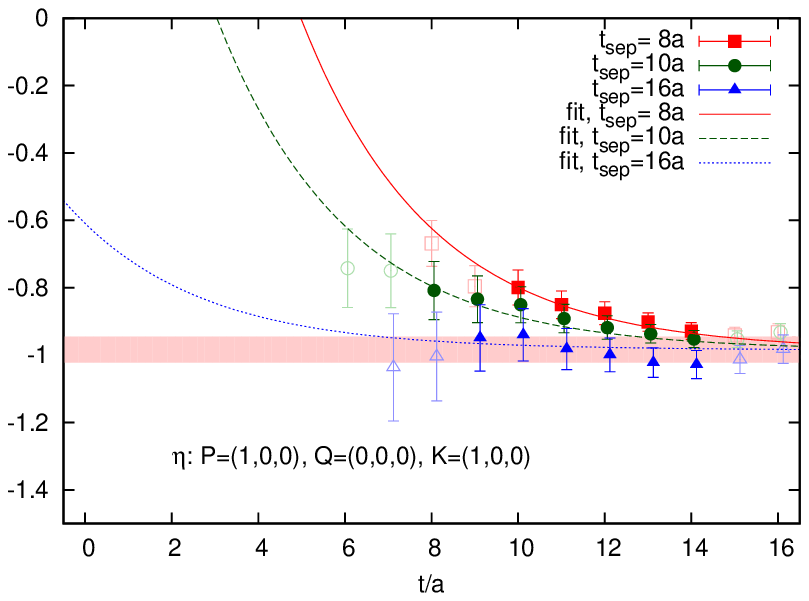}
\hfil
 \includegraphics{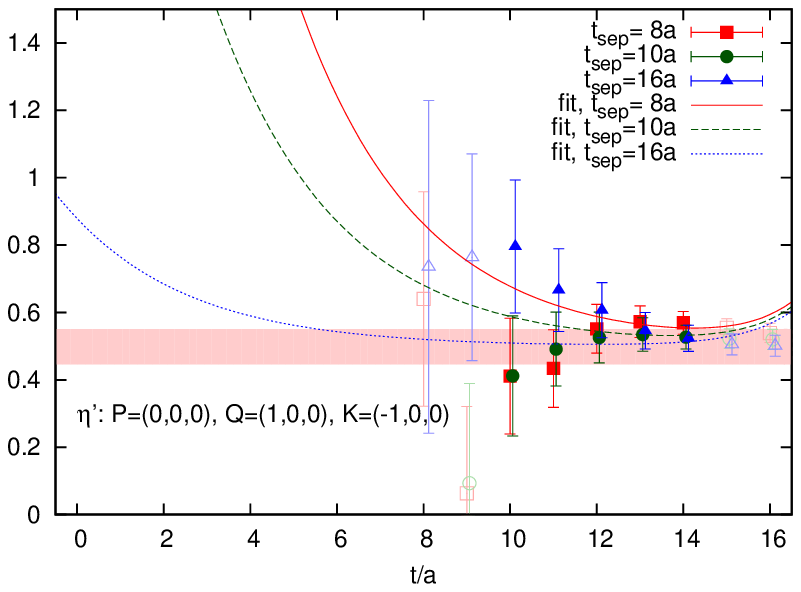}

\noindent
 \includegraphics{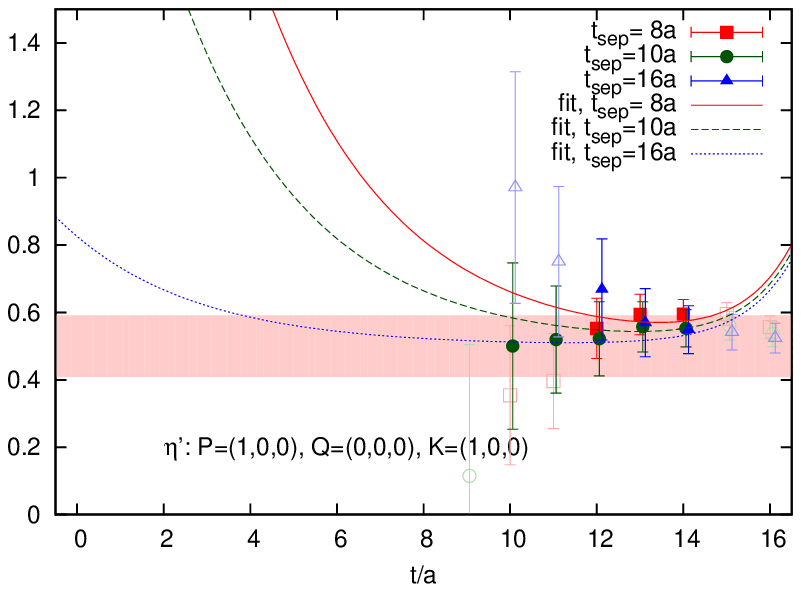}
\hfil
 \includegraphics{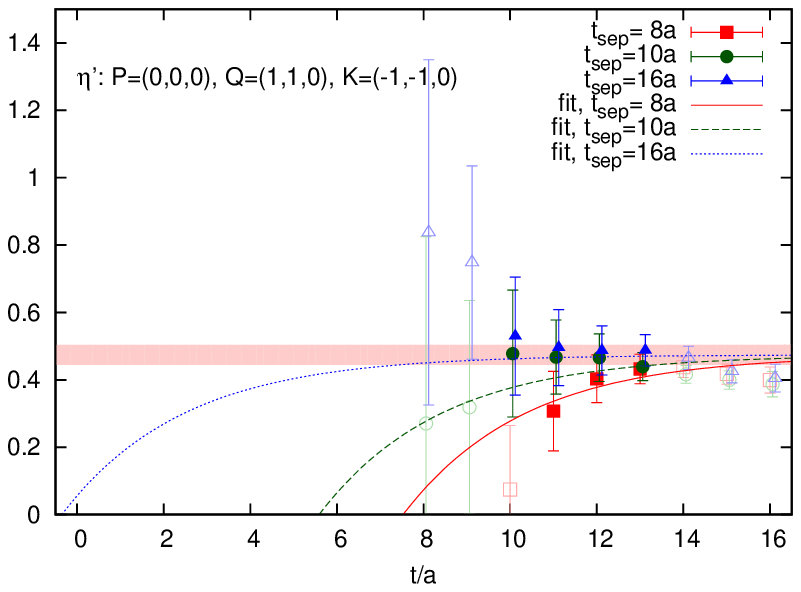}
\caption{Typical examples of fitting $R(t)$ to
  Eq.~(\protect\ref{eq:const_plus_excited}), to extract $\langle
  \eta^{(\prime)}(\V{k})|S(\V{0})|D_s(\V{p})\rangle$, for Set A.  The
  lower right plot depicts a fit to the first two terms of
  Eq.~(\protect\ref{eq:const_plus_excited}), while the others use all three
  terms. The $D_s$ meson is always located at $t/a=16$, while the
  $\eta$ or $\eta'$ is located at $t/a=8$ ($t_{\rm sep}/a=8$), $t/a=6$
  ($t_{\rm sep}/a=10$) and $t=0$ ($t_{\rm sep}/a=16$).  Data points with
open symbols were omitted from the fits. The red bands indicate
the values of the
  matrix elements obtained from the fit.
 The momenta are in the lattice units: 
 $(\V{P}, \V{Q}, \V{K}) = (\V{p}, \V{q}, \V{k}) \times L/(2\pi)$.
\label{fig:const_plus_excited}
}
\end{figure}

We generated three different data sets with $t_{\rm sep}/a=8,10,16$ and
fitted these simultaneously. For the $\eta$ at the SU(3) flavour
symmetric point, which has no disconnected contributions, we also generated
$t_{\rm sep}/a=24$ data. For some momentum combinations only a
subset of the available data was used in the fits, either
due to the data being too noisy (for $t_{\rm sep}/a=24$) or because
contributions from the second or higher excited states were significant
(for $t_{\rm sep}/a=8$).
Details of the chosen fit
ranges are listed in
Tables~\ref{tab:f0_eta_setA}--\ref{tab:f0_etap_setS} of
Appendix~\ref{app:data} and typical examples of the fits using
Eq.~(\ref{eq:const_plus_excited}) are shown in
Fig.~\ref{fig:const_plus_excited}.

Again, we used correlated fits and varied the fit region to
assess systematic uncertainties.
The changes of the fit parameter values were found to be well
within the statistical errors. The only exception was for the
three-point function involving the $\eta$ meson at the SU(3) flavour
symmetric point.  In this case, 
the statistical errors were small such that the
systematic uncertainties became relevant and we opted for
employing an uncorrelated fit and a fit range that
resulted in errors large enough to encompass the systematics.

\subsection{Results}
\label{sec:results}

\begin{figure}
 \noindent
 \hfil
 \includegraphics[width=0.48\linewidth]{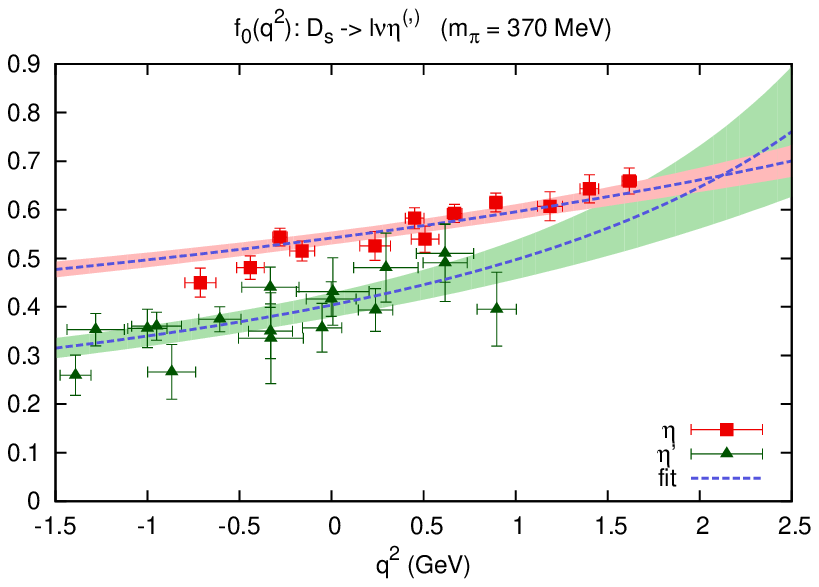}
 \hfil
 \includegraphics[width=0.48\linewidth]{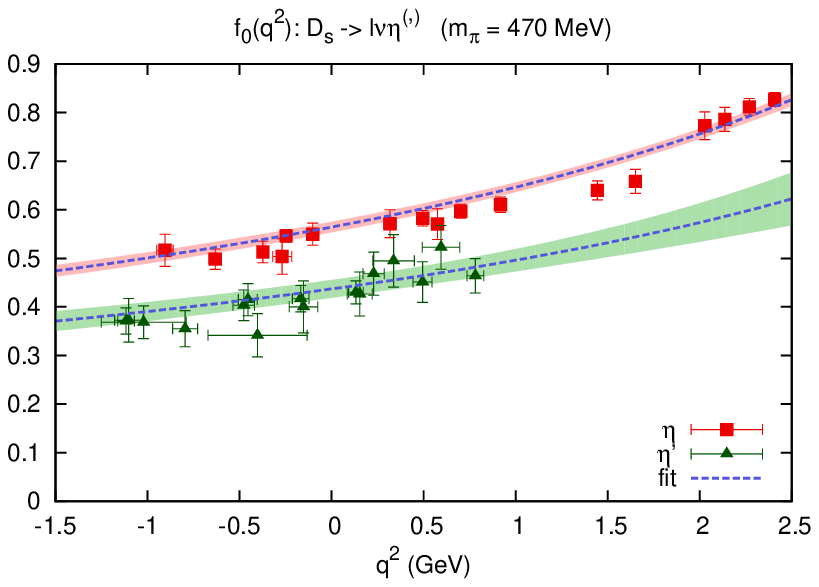}
\caption{The scalar form factor $f_0(q^2)$ for $D_s\to \eta^{(\prime)}
  \ell\bar{\nu}_{\ell}$.  The errors are statistical only and the dashed lines
  indicate the fits to the form factors using the parameterization
  $f_0(q^2)=f_0(0)/(1-bq^2)$. On the left are the results for $M_\pi\approx
  370\, {\rm MeV}$ (Set A) and on the right for $M_\pi\approx 470\, {\rm MeV}$ (Set S).}
\label{fig:f0}
\end{figure}

The results for the form factor, derived from the matrix elements
using Eq.~(\ref{eq:scalar-current}), are listed in
Tables~\ref{tab:f0_eta_setA}--\ref{tab:f0_etap_setS} for the momentum
ranges $\V{P}^2 \leq 4$ and $\V{K}^2 \leq 3$ in lattice units
($\V{P}=\V{p} \times L/(2\pi)$).
Note that we defined the four-momentum transfer $q^2$ as
\begin{align}
q^2 
 \equiv 
 \left(E_{D_s}(\V{p}) - E_{\eta^{(\prime)}}(\V{k})\right)^2
  - (\V{p}-\V{k})^2\,,
\end{align}
where the energies of the $D_s$ and $\eta^{(\prime)}$ states
are listed in Tables~\ref{tab:2pt_setA} and \ref{tab:2pt_setS}
of Appendix~\ref{app:data} and depicted in
Fig.~\ref{fig:dispersion}. The values at non-zero momenta were
determined directly, without using a dispersion relation. The
dependence of $f_0(q^2)$ on $q^2$ is shown
in Fig.~\ref{fig:f0}.

We used a one pole ansatz to
interpolate the data to $q^2=0$:
\begin{equation}
 f_0(q^2) = \frac{f_0(0)}{1-bq^2}\,.
\label{eq:simple-param}
\end{equation}
These curves are also shown in the figure.
The resulting values for $f_0(0)$ are listed in Table~\ref{tab:f0}.
The parameterization of Becirevi\'c and Kaidalov (BK)
\cite{Becirevic:1999kt} is frequently used in the literature too. For the
scalar form factor, this is essentially the same
parameterization as Eq.~(\ref{eq:simple-param}) but the location of
the pole is normalized with respect to the
vector meson mass $M_{D_s^*}$:\footnote{This
should not be confused with the excited state
of the pseudoscalar meson which we also denoted by a $*$ in
the previous subsection.}
\begin{equation}
 f_0(q^2) = \frac{f_0(0)}{1-x/\beta}
\end{equation}
with $x=q^2/M_{D_s^*}^2$.
The values of $\beta$ obtained from rescaling the parameter $b$
above are also listed in Table~\ref{tab:f0}.

A comparison can be made with the values derived from light cone QCD
sum rules (LCSRs) \cite{Offen:2013nma}, displayed in Table~\ref{tab:f0}, where
we assumed $f_+(q^2=0) =f_0(q^2=0)$.  Encouragingly, the
results are broadly consistent.
We find $f_0(q^2=0)$ is larger for the $\eta$ than for the $\eta'$,
independent of the quark mass, while for LCSRs the ordering cannot be
resolved due to the large error for the $\eta'$. 
The ratios of the form factors
$|f_+^{D_s \to \eta'}(0)|/|f_+^{D_s \to \eta}(0)| 
=|f_0^{D_s \to \eta'}(0)|/|f_0^{D_s \to \eta}(0)| $
are
\begin{align}
 &0.775(032) \quad \text{(Set S)}, &
 &0.746(046) \quad \text{(Set A)}, &
 &1.20(17) \quad \text{(LCSRs)}.
\label{eq:f0byf0}
\end{align}
A more detailed comparison would require an estimation of
the dominant systematic uncertainties. These uncertainties are
difficult to quantify in both studies, in the LCSRs case due to the
approximations made, while in our study since we have a single lattice
volume and lattice spacing. 
Considering our lightest pseudoscalar mass is around $370$~MeV
and $LM_\pi=3.3$, extending the
analysis to bigger volumes and smaller quark masses
is important.

\begin{table}
\caption{Parameters $f_0(0)$ and $b$
obtained from a fit  $f_0(q^2)=f_0(0)/(1-bq^2)$.
The coefficient $\beta$
  corresponds to an equivalent fit using the BK~\protect\cite{Becirevic:1999kt}
  parameterization.  The
  light cone QCD sum rule results (LCSRs) are taken from
  Ref.~\protect\cite{Offen:2013nma}.}
\label{tab:f0}
\begin{ruledtabular}
  \begin{tabular}{ccccc}
  Set & meson & $f_0(q^2=0)$ &  $b \, ({\rm GeV})^{-2} $  & $\beta$ \\
 \hline
  S   & $\eta$ & 0.564(11) & 0.127(06) & 1.70(08) \\
      & $\eta'$ &  0.437(18) & 0.119(23) & 1.81(35) \\
 \hline
  A   & $\eta$ &  0.542(13) & 0.090(14) & 2.35(36)\\
      & $\eta'$ & 0.404(25) & 0.188(32) & 1.13(19)\\
 \hline
 LCSRs (at $M_\pi^{\rm phys}$) 
  & $\eta$ &  0.432(33) & --- & --- \\
  & $\eta'$ & 0.520(80) & --- & ---
 \end{tabular}
\end{ruledtabular} 
\end{table}

\subsection{Outlook on phenomenology}
\label{sec:outlook}

The results given in the previous subsection do not allow for a direct
determination of the widths $\Gamma(D_s^- \to \eta e^- \bar\nu_e)$ and
$\Gamma(D_s^- \to \eta' e^- \bar\nu_e)$, since we computed $f_0(q^2)$ rather
than $f_+(q^2)$ and  used heavier-than-physical
pion masses.
Accordingly, a direct comparison to, for example,  the ratio
$\Gamma(D_s^- \to \eta' e^- \bar\nu_e)/\Gamma(D_s^- \to \eta e^- \bar\nu_e)=0.36(14)$, as
determined by the CLEO collaboration \cite{Yelton:2009aa}, is not yet possible.
However, invoking some model assumptions, a tentative comparison can
be made, albeit at the price of introducing an essentially unquantifiable
uncertainty.

We calculate the ratio
\begin{equation}
\frac%
{\Gamma(D_s^- \to \eta' e^- \bar\nu_e)}
{\Gamma(D_s^- \to \eta  e^- \bar\nu_e)}=
\frac%
{\int_0^{(M_{D_s}-M_{\eta'})^2} \lambda_{D_s,\eta'}^{3/2}(q^2) |f_+^{D_s\to\eta'}(q^2)|^2 dq^2}
{\int_0^{(M_{D_s}-M_{\eta })^2} \lambda_{D_s,\eta }^{3/2}(q^2) |f_+^{D_s\to\eta }(q^2)|^2 dq^2},
\label{eq:pheno_ratio}
\end{equation}
where $\lambda_{H,P}(x)$ is the 
heavy-light kinematic factor
\begin{equation}
\lambda_{H,P}(x)=\frac{1}{4M_H^2}\Big( (M_H^2+M_P^2-x)^2 - 4M_H^2M_P^2 \Big)
\;,
\end{equation}
by replacing $f_+(q^2)$ with the Ball-Zwicky
ansatz \cite{Ball:2004ye}
\begin{equation}
f_+^\mathrm{BZ}(q^2)=f_0(0)\,
\Big(
\frac{1}{1-q^2/M_{D_s^*}^2}+
\frac{rq^2/M_{D_s^*}^2}{(1-q^2/M_{D_s^*}^2)(1-\alpha q^2/M_{D_s^*}^2)}
\Big)\label{ballz}
\end{equation}
and using a chirally extrapolated value of our lattice results for
$f_0(0)$.  $M_{D_s^*},\alpha,r$ are taken from the literature. We
choose to compute the ratio rather than the individual decay rates
since systematics in the chiral extrapolation and the phenomenological
parameterisation of $f_+(q^2)$ partially cancel between the decay rates
for $\eta$ and $\eta^\prime$.

\begin{figure}
\noindent
\hfil
\includegraphics[height=0.35\linewidth]{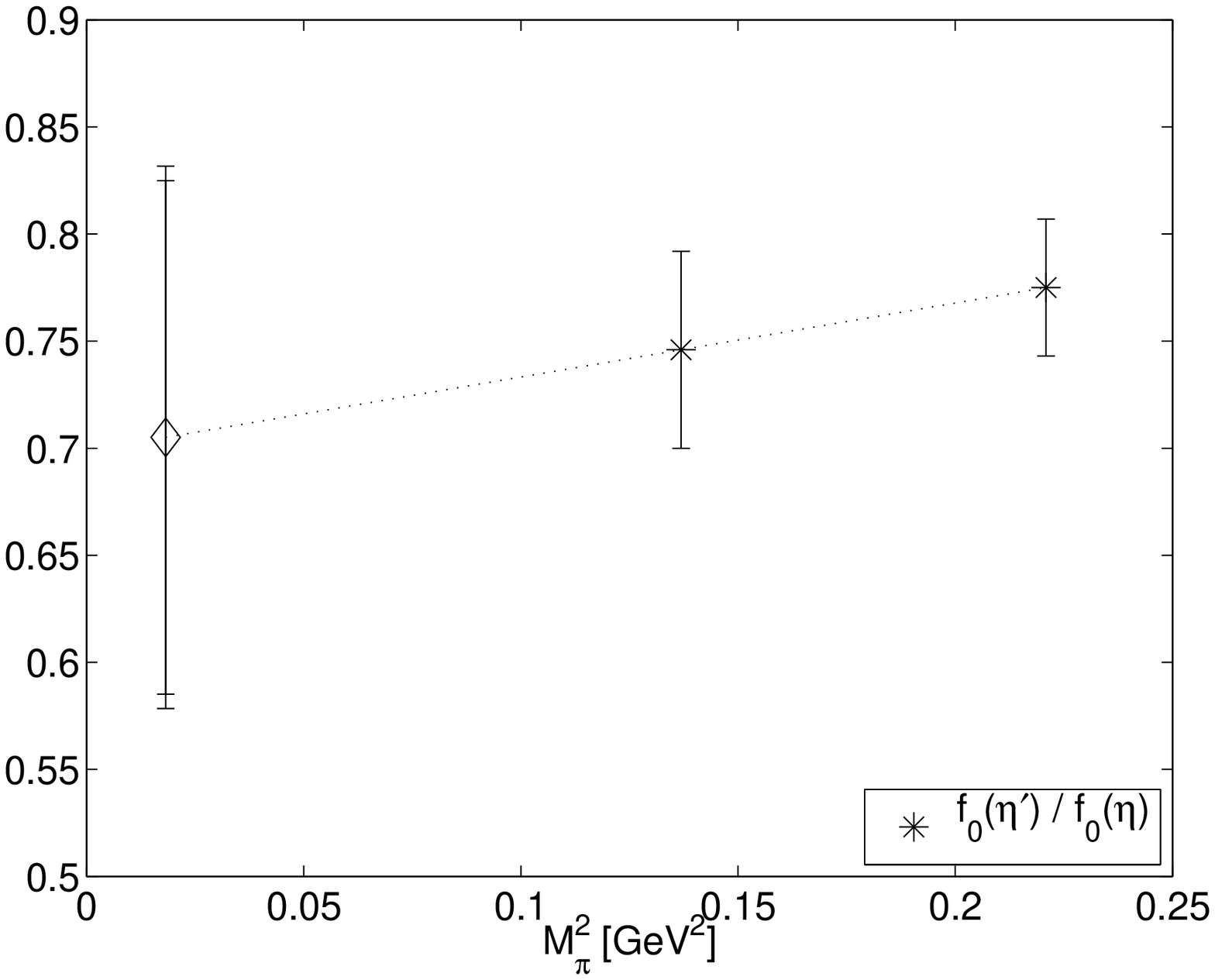}
\hfil
\includegraphics[height=0.35\linewidth]{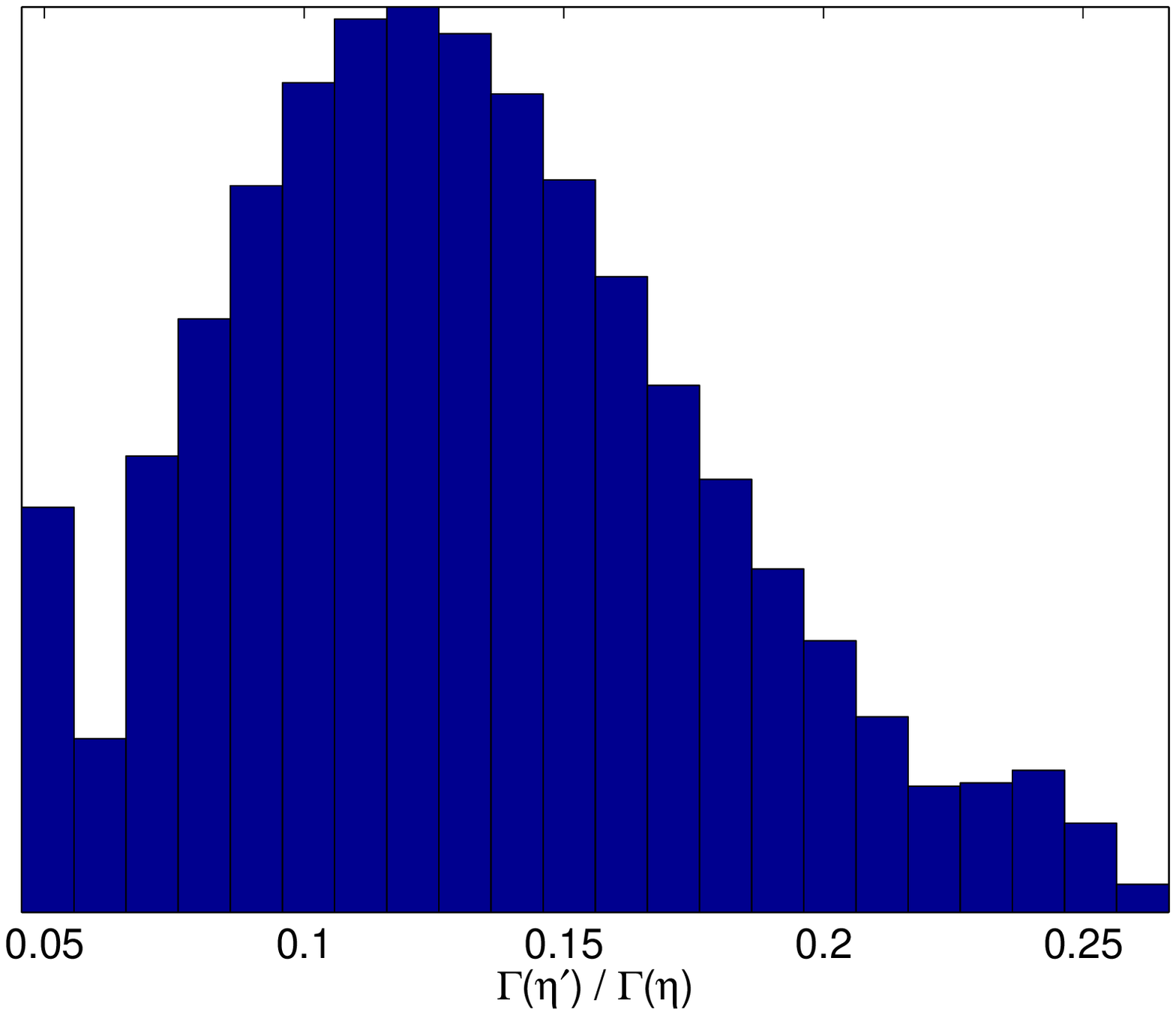}
\caption{\label{fig:pheno}
Left: Extrapolation of the ratio $f_0^{D_s\to\eta'}(0)/f_0^{D_s\to\eta}(0)$ to
the physical pion mass. The inner
error-bar of the extrapolated ratio is statistical, the outer one includes systematics~(see text for details).
Right: Histogram of the ratio
Eq.\,(\ref{eq:pheno_ratio}), varying the ratio $f_0^{D_s\to\eta'}(0)/f_0^{D_s\to\eta}(0)$
within its errors as well as the parameters $r$ and $\alpha$
within~Eq.~(\ref{ballz}) for the decays into $\eta$ and $\eta'$.}
\end{figure}

Using the above parameterisation only the ratio
$f_0^{D_s\to\eta'}(0)/f_0^{D_s\to\eta}(0)$ enters in Eq.~(\ref{eq:pheno_ratio}); thus we
extrapolate our two values for the ratio, given in
Eq.\,(\ref{eq:f0byf0}), linearly in $M_\pi^2$ to the physical mass
point, see Fig.~\ref{fig:pheno}; this yields
$f_0^{D_s\to\eta'}(0)/f_0^{D_s\to\eta}(0)=0.705(120)(041)$, where the
first uncertainty is statistical and the second one is systematic
(taken as the difference between the central values at
$M_\pi=370\,\mathrm{MeV}$ and $135\,\mathrm{MeV}$).  For $M_{D_s^*}$
we take the experimental value~\cite{Agashe:2014kda}, and for $\alpha$, $r$ we use
the central values determined in Ref.\,\cite{Offen:2013nma}, 
with 50\% uncertainties: $\alpha=0.252(126)$ and
$r=0.284(142)$.

We vary $\alpha$ and $r$ independently within the
form factors $f_+^{D_s\rightarrow\eta}$ and $f_+^{D_s\rightarrow\eta'}$ and
evaluate Eq.~(\ref{eq:pheno_ratio}) assuming Gaussian distributions of the five 
parameters $f_0^{D_s\to\eta'}(0)/f_0^{D_s\to\eta}(0)$, $r_\eta$, $\alpha_\eta$,
$r_{\eta'}$ and $\alpha_{\eta'}$ within the respective errors given above.
In the right panel of Fig.~\ref{fig:pheno} the resulting histogram
is shown. We find 
\begin{equation}
\frac{\Gamma(D_s^- \to \eta' e^- \bar\nu_e)}{\Gamma(D_s^- \to \eta  e^- \bar\nu_e)} =0.128^{+51}_{-42}\,,
\end{equation}
which deviates by 1.6$\sigma$ from the CLEO result.

Taking these numbers at face value would be premature.  We recall that
we used both a chiral extrapolation (our computations were performed
at $M_\pi=470\,\mathrm{MeV}$ and $M_\pi=370\,\mathrm{MeV}$) and a
model for the form factors $f_+^{D_s\to\eta}(q^2)$ and
$f_+^{D_s\to\eta'}(q^2)$.  Our result, however, demonstrates the
potential that a future lattice study of the form factors
$f_+^{D_s\to\eta}(q^2)$ and $f_+^{D_s\to\eta'}(q^2)$ will have.  A
study of these form factors, performed at the physical mass point,
will significantly reduce the errors, which at present are dominated
by systematics.  We recall that the present work is mainly a pilot
study to establish the feasibility of computations of form factors
involving quark line disconnected diagrams.

\section{Conclusions and discussion}
\label{sec:conclusions}
We calculated semileptonic decay form factors $f_0(q^2)$ 
for $D_s\to \eta \ell \bar{\nu}_{\ell}$ and $D_s \to \eta' \ell \bar{\nu}_{\ell}$ decays,
by means of numerical lattice simulation.
We included all disconnected fermion loop contributions.
Despite the statistically noisy and computationally expensive
disconnected part, we obtained the form factor at $q^2=0$
within statistical errors of less than 6\%.
The values at $q^2=0$ are
$|f_0^{D_s\to \eta}|=0.564(11)$ 
and
$|f_0^{D_s\to \eta'}|=0.437(18)$
at $M_\pi\approx 470\, {\rm MeV}$ and
$|f_0^{D_s\to \eta}|=0.542(13)$ 
and
$|f_0^{D_s\to \eta'}|=0.404(25)$
at $M_\pi\approx 370\, {\rm MeV}$,
where the errors are statistical only.
The masses of the $\eta$ and the $\eta'$  mesons
are $M_{\eta}=M_{\pi}=470.5(1.8)\,\mathrm{MeV}$
and $M_{\eta'}=1032(27)\, {\rm MeV}$ 
at $M_\pi\approx 470\, {\rm MeV}$,
and $M_{\eta}=542.8(6.2)$ and $M_{\eta'}=946(65)\, {\rm MeV}$ 
at $M_\pi\approx 370\, {\rm MeV}$, keeping $2M_K^2+M_{\pi}^2\propto m_s+2m_l$
approximately constant.
The mixing angle in the octet-singlet basis for 
the $M_\pi\approx 370\, {\rm MeV}$ case is 
$\theta_8=-10.9(1.5)_{\rm stat.}(0.5)_{\rm fit}^\circ$, 
$\theta_1=-5.5(1.5)_{\rm stat.}(1.2)_{\rm fit}^\circ$
and 
$\bar{\theta}=-7.7(0.9)_{\rm stat.}(0.8)_{\rm fit}^\circ$ 
in the parameterization Eq.~(\ref{eq:thetas}).
There is no mixing in the flavour
symmetric $M_\pi\approx 470\, {\rm MeV}$ case.
This means we have two different mixing angles, indicating
higher Fock state contributions. We are not yet able to
extrapolate the mixing angles, leading distribution amplitudes or
masses to the physical point, however, assuming
a monotonous dependence of the mixing angles on the light quark mass,
their absolute values should increase towards the physical point.

It is interesting to note that the disconnected fermion loop contribution
to $f_0^{D_s\to \eta'}$ is really significant.
In Fig.~\ref{fig:conn_vs_disconn} we saw that the relevant
three-point function contains a large contribution from the disconnected
diagram. This implies that the OZI suppressed gluonic contribution is not
suppressed in this decay mode due to the chiral anomaly, as is
also indicated by the fact that singlet and octet $\eta'$ distribution
amplitudes cannot be parameterized by a single angle, relative to
the octet-singlet basis.

We calculated the scalar form factor $f_0(q^2)$ which does not require
knowledge of the renormalization constants. In order to compare with
experiment, however, the vector form factor $f_+(q^2)$ is more
relevant, since, in the massless lepton limit, only $f_+(q^2)$
contributes to the decay width.  Technically, a computation of
$f_+(q^2)$ is of a similar level of complexity as the present study
and we plan to pursue this in the near future.  Finally, we remark
that this work is an exploratory study and the quark masses we used
are not yet physical. Having verified that computations of
disconnected contributions to form factors are feasible, lighter pion
masses and larger volumes will be simulated, also extending the
present study to decays with the $\phi$ in the final state.

\acknowledgments

We thank our collaborators within QCDSF who generated the $N_f=2+1$
ensembles analyzed here.  
We also thank the International Lattice DataGrid.
We thank Vladimir Braun, Benjamin Gl\"a{\ss}le, Meinulf G\"ockeler,
Johannes Najjar and
Paula P\'erez-Rubio for their help, useful comments and discussions.

This work was supported by the DFG (SFB/TRR 55) and the EU (ITN
STRONGnet). The CHROMA software suite \cite{Edwards:2004sx} was used
extensively along with the
locally deflated domain decomposition solver
implementation of openQCD~\cite{luscherweb}. We benefited from time granted
by PRACE (project 2012071240) on Fermi at CINECA, Bologna, as well as
the Athene HPC and iDataCool clusters at the University of Regensburg.
I. K. thanks the Ministry of Science and Technology in
Taiwan (the grand 103-2811-M-009-014) and the hospitality
of NCTS-north.

\appendix

\section{Details of the estimation of disconnected loops}
\label{app:disconn_loop}

In this Appendix we explain the methods implemented to calculate the
disconnected loop given in Eq.~(\ref{eq:1pt-def}). For convenience we
restate the equation as
\begin{align}
 C_{\rm 1pt}(t,\V{p}; \V{x}_0)
 &=  \sum_\V{x}  \exp(i\V{p}\cdot(\V{x}-\V{x}_0)) C_{\rm 1pt}(t,\V{x})\,,\\
C_{\rm 1pt}(t,\V{x})
 & =  \tr \left[
     \sum_{\V{x}',\, \V{x}''}
    \Gamma \phi(\V{x},\V{x}'') M^{-1}(t,\V{x}''; t,\V{x}')
    \phi(\V{x}',\V{x})
     \right]\, .
 \label{eq:1pt-def_second}
\end{align}
The all-to-all propagator, $M^{-1}(t,\V{x}''; t,\V{x}')$, is computed
using low mode deflation combined with stochastic estimation. This
involves 
calculating $n_{\rm low}$ (exact) low eigenmodes (in absolute magnitude)
of the Hermitian Dirac 
operator $Q= \gamma_5 M$:
\begin{equation}
 Q|\lambda_i\rangle = \lambda_i |\lambda_i\rangle\,,
\end{equation}
where the $\lambda_i$ are real.
The low mode contribution to $M^{-1}$ is given by
\begin{equation}
 M^{-1}|_{\rm low} 
= \sum_{i}\frac{1}{\lambda_i} |\lambda_i\rangle \langle \lambda_i| \gamma_5\,.
\end{equation}
For small quark masses the low modes give the most singular directions
of $M^{-1}$ and the higher
mode contributions $M^{-1}|_{\rm high}=M^{-1}- M^{-1}|_{\rm low}$ become small.
These higher modes are estimated stochastically
using $\frac{1}{\sqrt{2}}(\mathbb{Z}_2 +i\mathbb{Z}_2)$ noise vectors,
which approximately span a complete set
\begin{equation}
 \frac{1}{n_{\rm stoch}}\sum_{s=1}^{n_{\rm stoch}}
  |\eta_s\rangle \langle \eta_s|
 = 1 + O\left( \frac{1}{\sqrt{n_{\rm stoch}}} \right)\,.
\end{equation}
We have
\begin{equation}
 M^{-1}|_{\rm high}
=\frac{1}{n_{\rm stoch}}\sum_{s=1}^{n_{\rm stoch}}
 M^{-1}|\tilde{\eta}_s\rangle \langle \eta_s|\,,
\end{equation}
where 
\begin{equation}
 |\tilde{\eta}_s\rangle 
 = \gamma_5 \left( 1- \sum_{i=1}^{n_{\rm  low}}
	     |\lambda_i\rangle \langle\lambda_i|
	    \right)\gamma_5 |\eta_s\rangle
 \label{eq:high_noise}
\end{equation}
is the source vector projected onto the subspace of the higher modes.

Stochastic estimation introduces additional (possibly dominant) noise on top of
the gauge noise and this needs to be reduced. We implemented a number of
techniques to achieve this:
\begin{enumerate}
\item Time and spin partitioning~\cite{Bernardson:1993he}.  The
  stochastic sources where given non-zero values only on every 4th
  timeslice and for a single spin index. To reconstruct the full
  propagator at every timeslice requires $4(\text{spin}) \times
  4(\text{time}) = 16$ inversions.
\item Hopping parameter acceleration (HPA)~\cite{Thron:1997iy}.
A Wilson-type Dirac operator
\begin{equation}
 M=\frac{1}{2\kappa} (1-\kappa D)
\end{equation}
satisfies the identity:
\begin{equation}
 M^{-1} 
  = 2\kappa \sum_{i=0}^\infty (\kappa D)^i
  =  2\kappa \sum_{i=0}^{n-1} (\kappa D)^i + (\kappa D)^n  M^{-1} 
\end{equation}
for any integer $n\geq 0$.
When this expression is inserted into Eq.~(\ref{eq:1pt-def_second})
the first $n$ terms may be zero, where the value of $n$ depends on
$\Gamma$ and the form of the Dirac operator. With stochastic
estimation of $M^{-1}$ these terms will only contribute to the noise
and can be omitted, giving $(\kappa D)^n M^{-1}$ as an improved
estimate of $M^{-1}$.

Combining this with low mode deflation we have
\begin{equation}
 M^{-1}
  =  \sum_{i=1}^{n_{\rm  low}}
      \frac{1}{\lambda_i} 
      (\kappa D)^n |\lambda_i\rangle
       \langle\lambda_i|\gamma_5
    + \frac{1}{n_{\rm stoch}}\sum_{s=1}^{n_{\rm stoch}}
       (\kappa D)^n M^{-1} |\tilde{\eta}_s \rangle
      \langle \eta_s|\,.
\label{eq:stoch_estimation}
\end{equation}
For the clover action and $\Gamma=\gamma_5$ we can use $n=2$.

\item The truncated solver method (TSM)~\cite{Bali:2009hu}.  This
  method involves truncating the solver after a few iterations. The
  (hopefully small) correction to this truncation is calculated using
  a smaller number of stochastic estimates:
\begin{equation}
\frac{1}{n_{\rm stoch}}\sum_{s=1}^{n_{\rm stoch}} M^{-1}|\tilde{\eta}_s\rangle \langle \eta_s| \mapsto
\frac{1}{N_1}\sum_{s=1}^{N_1} M^{-1}_{\mathrm{trunc}}|\tilde{\eta}_s\rangle\langle \eta_s|+
\frac{1}{N_2}\sum_{s=N_1+1}^{N_1+N_2} (M^{-1}-M^{-1}_{\mathrm{trunc}})|\tilde{\eta}_s\rangle\langle \eta_s|\,,
\label{eq:tsm}
\end{equation}
where $N_2<N_1$.  The truncated part is calculated with a CG solver,
while for $M^{-1}|\tilde{\eta}_s\rangle$ we use the domain decomposition solver
implementation of Ref.~\cite{luscherweb}. 
To obtain the full expression for $M^{-1}$
using HPA and low mode deflation one substitutes Eq.~(\ref{eq:tsm}) into Eq.~(\ref{eq:stoch_estimation}).
\end{enumerate}

The parameters for the various techniques are chosen so that the
stochastic error is minimized for fixed computational cost, see
Ref.~\cite{Bali:2011yx} for details.  Our optimal choices are listed in
Table~\ref{tab:disconn_parameters}.  We found the HPA to be the most
cost efficient noise reduction technique for our problem.  The TSM
only provided a slight improvement, due to the use of smeared loops.
In general, the advantage of using the TSM will also depend on the
efficiency of the solver.

Finally, we note that due to parity and charge conjugation
considerations the disconnected loop in position space, 
$C_{\rm 1pt}(t,\V{x})$, is real for $\Gamma=\gamma_5$.
This means the imaginary part of our stochastic estimation of 
$C_{\rm 1pt}(t,\V{x})$ only contributes to the noise
and we can set it to zero.  

\begin{table}
\caption{Parameters for the estimation of the disconnected loop.  If
  the TSM is used, $n_{\rm stoch}$ stands for $N_1+N_2$, where $N_1$
  and $N_2$ are the numbers of stochastic estimates used for
 the truncated part and to estimate the bias, respectively. Note that due to
the  use of spin and time dilution, each stochastic estimation requires
  $16$ inversions of the noise vector.}
\label{tab:disconn_parameters}
\begin{ruledtabular}
  \begin{tabular}{ccccc}
  Set & quark & $n_{\rm low}$ & $n_{\rm stoch}$ & TSM \\
 \hline
 S  & $l$, $s$ & $24$ & $10+3$ & truncated after 150 CG iterations \\
 \hline
 A  & $l$    & $40$ & $24+8$ & truncated after 120 CG iterations \\
    & $s$   & $40$ & $48$   & (without TSM)
 \end{tabular}
\end{ruledtabular}
\end{table}

\section{Two-point functions}
\label{app:2pt}

In the octet-singlet basis, 
we need the following two-point functions:
\begin{align}
 \langle \mathcal{O}_8 (t;\V{p}) \mathcal{O}^\dagger_8(0)\rangle
 &= \frac{1}{3}\left\langle(
     C_{ll} +C_{ss}
      - 2D_{ll}- 2D_{ss} + 2D_{ls} + 2D_{sl}
  )\right\rangle,
 \label{eq:eta8-eta8}
 \\
 \langle \mathcal{O}_1 (t;\V{p}) \mathcal{O}^\dagger_1(0)\rangle
 &= \frac{1}{3}\left\langle(
     2C_{ll} +C_{ss}
      - 4D_{ll}- D_{ss} - 2D_{ls} - 2D_{sl}
  )\right\rangle, 
 \label{eq:eta1-eta1}
\\
 \langle \mathcal{O}_1 (t;\V{p}) \mathcal{O}^\dagger_8(0)\rangle
 &= \frac{\sqrt{2}}{3}\left\langle(
     C_{ll} - C_{ss}
      - 2D_{ll} + D_{ss} + 2D_{ls} - D_{sl}
  )\right\rangle, 
 \label{eq:eta8-eta1}
\\
 \langle \mathcal{O}_8 (t;\V{p}) \mathcal{O}^\dagger_1(0)\rangle
 &= \langle \mathcal{O}_1 (t;\V{p}) \mathcal{O}^\dagger_8(0)\rangle^*\,,
 \label{eq:eta1-eta8}
\end{align}
where $C_{aa}=C_{aa}(t,\V{p})$ is a connected two-point function of
quark flavour $a=l,s$ and $D_{ab}$ is the disconnected two-point
function of quark flavours $a$ and $b$:
\begin{equation}
D_{ab} (t, \V{p}) 
 = \frac{a^4}{V_4} \sum_{t_0/a=0}^{T/a-1}
   C_{\rm 1pt}^{a}(t+t_0, \V{p}) C_{\rm 1pt}^{b}(t_0, -\V{p}),
 \label{eq:disconn2pt}
\end{equation}
where $T$ is the temporal lattice size, $V_4$ is the four-volume
and $C_{\rm 1pt}^a(t,\V{p})$ is the disconnected fermion loop,
Eq.~(\ref{eq:1pt-def}), for quark flavour $a$.

The calculation of $C_{\rm 1pt}^a(t,\V{p})$ is detailed
in Appendix~\ref{app:disconn_loop}.
For the connected two-point function, we implemented low mode
averaging (LMA)~\cite{DeGrand:2004qw,Giusti:2004yp} 
reusing the eigenmodes computed for the evaluation of the disconnected loop.
As discussed in Ref.~\cite{Bali:2010se}, 
LMA works very efficiently for pseudoscalar meson two-point functions.
We used LMA for both the connected light-light ($C_{ll}$)
and strange-strange ($C_{ss}$) two-point functions.  

A connected two-point function with LMA is given by
\begin{equation}
 C_{\rm LMA}^{\rm 2pt}(t,\V{p})
 =  C_{\rm pa}^{\rm 2pt}(t,\V{p}; x_0)  - C_{\rm low,pa}^{\rm 2pt}(t,\V{p}; x_0)
    + C_{\rm low}^{\rm 2pt}(t,\V{p})\,,\label{eq:lowmodeav}
\end{equation}
where $C_{\rm pa}^{\rm 2pt}(t, \V{p}; x_0)$ is the standard
point-to-all two-point function, calculated with a single source point
at $x_0=(t_0,\V{x}_0)$.  For simplicity, we have suppressed the quark
flavour index and, initially, do not consider quark smearing.  In
Eq.~(\ref{eq:lowmodeav}), the low mode contribution to the
point-to-all two-point function,
\begin{equation}
 C_{\rm low, pa}^{\rm 2pt}(t, \V{p} ; x_0)
 = \sum_{\V{x}} \exp(i \V{p}\cdot\V{x})
    \sum_{i,j=1}^{n_{\rm low}}\frac{1}{\lambda_i \lambda_j}
    \langle \lambda_i(x_0) | \gamma_5 \Gamma^\dagger
   | \lambda_j(x_0)\rangle \langle \lambda_j(x+x_0) |
   \gamma_5 \Gamma
   | \lambda_i(x+x_0) \rangle\,, 
 \label{eq:2pt_low_pa}
\end{equation}
where $x=(t,\V{x})$ and $\Gamma=\gamma_5$ 
at the source and sink, is subtracted and replaced by the low mode contribution 
averaged over all lattice points:
\begin{equation}
 C_{\rm low}^{\rm 2pt}(t, \V{p})
 =    \frac{a^4}{V_4}\sum_{x_0} C_{\rm low, pa}^{\rm 2pt}(t,\V{p}; x_0)\,.
 \label{eq:2pt_low}
\end{equation}
Smearing the quarks is implemented by replacing the eigenvectors
$|\lambda_i\rangle $ in Eq.~(\ref{eq:2pt_low_pa}) with smeared vectors,
$\phi |\lambda_i\rangle$, for a smearing function $\phi$.

Finally, we averaged over forward and backward propagating two-point
functions, as well as rotationally equivalent momentum combinations.

\section{Three-point functions}
\label{app:3pt}

The three-point function we need to determine is
\begin{align}
 \left\langle
 C_{\rm 3pt}^{D_s\to \eta^{(\prime)}}(t,\V{p},\V{k};t_{\rm sep},x_0)
 \right\rangle
&= 
\langle 0| \mathcal{O}_{\eta^{(\prime)}}(\V{k},t_{\rm sep} + t_0) S(\V{q},t+t_0)
     \mathcal{O}_{D_s}^\dagger(t_0,\V{x}_0) |0 \rangle
 \nonumber\\
&= 
 \sum_{\V{x},\V{y}} e^{i\V{k}\cdot\V{x}} e^{i\V{q}\cdot\V{y}}
\langle 0|
 \mathcal{O}_{\eta^{(\prime)}}(t_{\rm sep} + t_0,\V{x}+\V{x}_0) 
  S(t+t_0,\V{y}+\V{x}_0)
     \mathcal{O}_{D_s}^\dagger(t_0,\V{x}_0) |0 \rangle\,,
 \label{eq:3pt_def}
\end{align}
where $\mathcal{O}_{\eta^{(\prime)}}$, $\mathcal{O}_{D_s}$ are the
interpolators, $S$ is the local scalar current
and $\V{p}=\V{q}+\V{k}$.
The interpolators for $\eta$ and $\eta'$ are obtained from
Eq.~(\ref{eq:physical_interpolator}), by solving the generalized
eigenvalue problem for each $\V{k}$.  For $\V{k}=\V{0}$, we used the
improved mixing angles $\theta$ and $\theta'$ as discussed 
in Sec.~\ref{sec:eta-etap-states}.

We need both connected and disconnected contributions to calculate the
three-point function Eq.~(\ref{eq:3pt_def}), see Fig.~\ref{fig:3pt}.
For the connected part, we used the stochastic method
detailed in Ref.~\cite{Evans:2010tg}.  This approach allows us to access many 
momentum combinations at a lower computational cost compared to the standard
sequential source method. We compute all rotationally equivalent momentum
combinations average these.

The disconnected part is obtained from combining a connected
charm-strange two-point function, $ C_{\rm cs}(t,\V{q};x_0)$, with a
one-point quark loop of flavour $a$:
\begin{equation}
 C_{\rm disc}^{{\rm 3pt},a} (t, t_{\rm sep},\V{p},\V{k};x_0)
 = C_{\rm 1pt}^a(t_0+t_{\rm sep}, \V{k};\V{x}_0) C_{\rm cs}(t,\V{q};x_0)\,,
\end{equation}
and
\begin{equation}
 C_{\rm cs}(t,\V{q};x_0) =
\sum_{\V{x}}\exp(i\V{q}\cdot\V{x})
\tr\left[
    M^{-1}_c(x+x_0;x_0) \gamma_5 M^{-1}_s(x_0; x+x_0)
 \right]\,.
\end{equation}
Note that the charm-strange
two-point function has a pseudoscalar source and a scalar sink.
The one-point loop is calculated as described in Appendix~\ref{app:disconn_loop}.

We employ low mode averaging in a similar way to that used
for the computation of the connected two-point
function in Appendix~\ref{app:2pt}, by averaging the low mode contributions 
to $C_{\rm dics}^{{\rm 3pt},a}$ over $x_0$:
\begin{align}
 C_{\rm disc, LMA}^{{\rm 3pt}, a}(t,t_{\rm sep},\V{p},\V{k};x_0)
 &=  C_{\rm 1pt}^a(t_0+t_{\rm sep}, \V{k};\V{x}_0) \left[
      C_{\rm cs}(t,\V{q};x_0) - C_{\rm low, pa}(t,\V{q};x_0)\right]
 \nonumber\\
 &\quad
 + \frac{1}{N_y}\sum_{y}
      C_{\rm 1pt}^a(t_0+t_{\rm sep}+t_y, \V{k}; \V{x}_0 + \V{y}) 
    C_{\rm low, pa}(t,\V{q};x_0+y))\,,
\end{align}
where
\begin{equation}
 C_{\rm low, pa}(t,\V{q}; x_0)
 \equiv \sum_\V{x} \exp(i\V{p}\cdot\V{x})
  \sum_{i=1}^{n_{\rm low}} \frac{1}{\lambda_i} \tr\bigl[
  M^{-1}_c(x+x_0;x_0) \gamma_5
   |\lambda_i(x_0)\rangle \langle \lambda_i (x+x_0)|
\bigr]
\end{equation}
and
\begin{equation}
 C_{\rm 1pt}^a(t_0+t_{\rm sep}+t_y, \V{k}; \V{x}_0+\V{y})
 =  C_{\rm 1pt}^a(t_0+t_{\rm sep}+t_y, \V{k};\V{x}_0) \exp(-i\V{k}\cdot\V{y}).
\end{equation}
The eigenvalues, $\lambda_i$, and eigenvectors, $|\lambda_i\rangle$, are
computed for the strange quark.
We average over $N_y=4^3< V_4/a^4$ source points only, due to the
computational cost of calculating the charm quark propagator,
$M^{-1}_c$, for each source. We employ
the subset $y=(t_y,\V{y})$ with $t_y=0$ and $\V{y}=(n_1,n_2,n_3)L/(4a)$
with $n_i=0,1,2,3$.

\begin{figure}
 \includegraphics{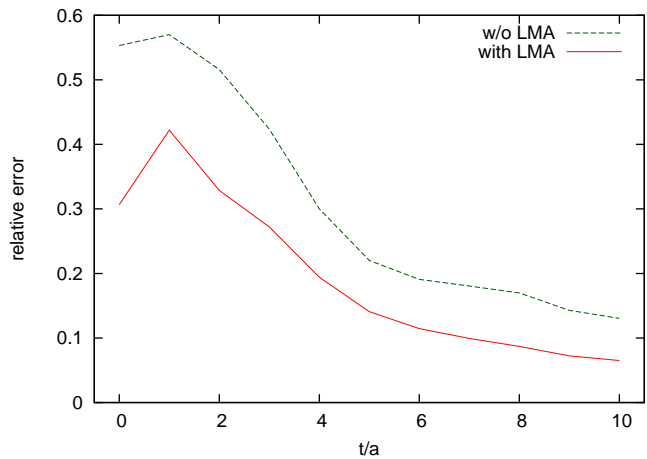}
 \caption{The relative 
errors of the disconnected three-point function
$\langle C_{\rm disc}^{{\rm 3pt}, l}(t,t_{\rm sep}, \V{p},\V{k})\rangle$ with
and without low mode averaging for Set S.
The sink-source separation is $t_{\rm sep}=10a$,
the $D_s$ meson is located at $t=0$ with lattice
momentum $\V{P}=(1,0,0)$ 
and the disconnected light-quark loop is
located at $t/a=10$ with momentum $\V{K}=(1,0,0)$. 
}
 \label{fig:LMA3pt}
\end{figure}

Finally, we averaged over rotationally equivalent momentum combinations,
and averaged over $+t_{\rm sep}$ and $-t_{\rm sep}$
for each value of sink-source separation $|t_{\rm sep}|$.
Fig.~\ref{fig:LMA3pt} shows a typical example of a comparison of
the relative error of the 
disconnected three-point 
function with and without low mode averaging.
The figure illustrates that LMA reduces the error significantly.

\section{Data}
\label{app:data}

The fitted values of two-point functions are listed in 
Tables~\ref{tab:2pt_setS} and \ref{tab:2pt_setA}.
The values of the scalar form factor $f_0(q^2)$ at each $q^2$ 
are listed in Tables~\ref{tab:f0_eta_setA}, \ref{tab:f0_etap_setA},
\ref{tab:f0_eta_setS}
and \ref{tab:f0_etap_setS}.

\begin{table}

\caption{
The ground state energies $E$ and energy gaps $\Delta E$ to
the first excited state for zero and finite momenta for
Set A. $\Delta E$ was obtained using Eq.~(\protect\ref{eq:def_DeltaE}).
The mass of the $\eta'$ meson is determined applying the
improved method (see Sec.~\protect\ref{sec:eta-etap-states}).
Also listed is the mass obtained by using the lattice dispersion relation
Eq.~(\protect\ref{eq:lattice_dispersion}) (lat.\ disp.).}
\label{tab:2pt_setA}
\begin{ruledtabular}
\begin{tabular}{cccccccc}

& & \multicolumn{3}{c}{ground state} & \multicolumn{3}{c}{excited state} \\
& $\V{p}\times L/(2\pi) $&
 fit range & $aE$ & $\chi^2/{\rm d.o.f}$ &  fit range &  $ a\Delta E$ & $\chi^2/{\rm d.o.f}$ \\
\hline
    &
   $(0,0,0)$ & 
   12--24 &
   $ 0.141 (01) $ &
   $ 1.08 $ & 
   --- &
   --- & 
   --- 
\\
    $\pi $ &
   $(1,0,0)$ & 
   9--17 &
   $ 0.296 (03) $ &
   $ 0.57 $ & 
   --- &
   --- & 
   --- 
\\
    &
   $(1,1,0)$ & 
   8--17 &
   $ 0.409 (18) $ &
   $ 0.78 $ & 
   --- &
   --- & 
   --- 
\\
    &
   $(1,1,1)$ & 
   6--11 &
   $ 0.522 (82) $ &
   $ 0.40 $ & 
   --- &
   --- & 
   --- 
\\
    &
   (lat. disp.)* & 
   --- &
   $ 0.141 (01) $ &
   $ 0.40 $ & 
   --- &
   --- & 
   --- 
\\
\hline
    &
   $(0,0,0)$ & 
   6--24 &
   $ 0.207 (03) $ &
   $ 0.73 $ & 
   2--4 &
   $ 0.638 (91) $ & 
   $ 0.50 $ 
\\
    $\eta $ &
   $(1,0,0)$ & 
   9--23 &
   $ 0.328 (05) $ &
   $ 0.57 $ & 
   2--7 &
   $ 0.437 (43) $ & 
   $ 0.16 $ 
\\
    &
   $(1,1,0)$ & 
   8--14 &
   $ 0.438 (15) $ &
   $ 0.71 $ & 
   2--4 &
   $ 0.645 (219) $ & 
   $ 0.02 $ 
\\
    &
   $(1,1,1)$** & 
   5--11 &
   $ 0.577 (45) $ &
   $ 0.26 $ & 
   --- &
   --- & 
   --- 
\\
    &
   (lat. disp.) & 
   --- &
   $ 0.206 (02) $ &
   $ 1.33 $ & 
   --- &
   --- & 
   --- 
\\
\hline
    &
   $(0,0,0)$ & 
   7--11 &
   $ 0.309 (38) $ &
   $ 0.43 $ & 
   2--4 &
   $ 0.417 (105) $ & 
   $ 0.03 $ 
\\
    $\eta' $ &
   $(1,0,0)$ & 
   6--11 &
   $ 0.452 (22) $ &
   $ 0.06 $ & 
   2--4 &
   $ 0.815 (716) $ & 
   $ 0.21 $ 
\\
    &
   $(1,1,0)$ & 
   6--10 &
   $ 0.552 (40) $ &
   $ 1.17 $ & 
   --- &
   --- & 
   --- 
\\
    &
   $(1,1,1)$** & 
   4--7 &
   $ 0.715 (81) $ &
   $ 0.30 $ & 
   --- &
   --- & 
   --- 
\\
    &
   (lat. disp.) & 
   --- &
   $ 0.360 (25) $ &
   $ 1.77 $ & 
   --- &
   --- & 
   --- 
\\
\hline
    &
   $(0,0,0)$ & 
   11--21 &
   $ 0.775 (02) $ &
   $ 0.15 $ & 
   2--6 &
   $ 0.356 (98) $ & 
   $ 0.45 $ 
\\
    $D_s $ &
   $(1,0,0)$ & 
   11--21 &
   $ 0.814 (03) $ &
   $ 0.26 $ & 
   2--6 &
   $ 0.333 (68) $ & 
   $ 0.47 $ 
\\
    &
   $(1,1,0)$ & 
   11--20 &
   $ 0.851 (05) $ &
   $ 0.58 $ & 
   2--6 &
   $ 0.324 (62) $ & 
   $ 0.49 $ 
\\
    &
   $(1,1,1)$ & 
   11--21 &
   $ 0.886 (07) $ &
   $ 0.93 $ & 
   2--6 &
   $ 0.321 (64) $ & 
   $ 0.52 $ 
\\
    &
   $(2,0,0)$ & 
   11--20 &
   $ 0.924 (10) $ &
   $ 0.18 $ & 
   2--6 &
   $ 0.371 (116) $ & 
   $ 0.37 $ 
\\
    &
   $(2,1,0)$ & 
   11--20 &
   $ 0.959 (15) $ &
   $ 0.42 $ & 
   2--6 &
   $ 0.396 (154) $ & 
   $ 0.39 $ 
\\
    &
   (lat. disp.)* & 
   --- &
   $ 0.774 (02) $ &
   $ 0.41 $ & 
   --- &
   --- & 
   --- 
\\
\end{tabular}
\hspace*{\fill} *: Not used.\\
\hspace*{\fill} **: Not included in the dispersion relation.%
\end{ruledtabular}
\end{table}

\begin{table}
\caption{The same as Table~\protect\ref{tab:2pt_setA} for Set S.}
\label{tab:2pt_setS}
\begin{ruledtabular}
\begin{tabular}{cccccccc}

& & \multicolumn{3}{c}{ground state} & \multicolumn{3}{c}{excited state} \\
& $\V{p}\times L/(2\pi)$ &
 fit range & $aE$ & $\chi^2/{\rm d.o.f}$ &  fit range &  $ a\Delta E$ & $\chi^2/{\rm d.o.f}$ \\
\hline
    &
   $(0,0,0)$ & 
   14--24 &
   $ 0.179 (01) $ &
   $ 0.55 $ & 
   5--12 &
   $ 0.359 (56) $ & 
   $ 0.79 $ 
\\
    $\eta  $ &
   $(1,0,0)$ & 
   8--22 &
   $ 0.320 (02) $ &
   $ 1.37 $ & 
   2--5 &
   $ 0.574 (34) $ & 
   $ 0.72 $ 
\\
  $(=\pi)$
    &
   $(1,1,0)$ & 
   8--18 &
   $ 0.420 (05) $ &
   $ 1.09 $ & 
   2--4 &
   $ 0.590 (76) $ & 
   $ 0.15 $ 
\\
    &
   $(1,1,1)$ & 
   6--15 &
   $ 0.500 (12) $ &
   $ 0.64 $ & 
   2--5 &
   $ 0.648 (101) $ & 
   $ 0.68 $ 
\\
    &
   $(2,0,0)$** & 
   4--10 &
   $ 0.626 (21) $ &
   $ 0.22 $ & 
   --- &
   --- & 
   --- 
\\
    &
   (lat. disp.)* & 
   --- &
   $ 0.179 (01) $ &
   $ 5.44 $ & 
   --- &
   --- & 
   --- 
\\
\hline
    &
   $(0,0,0)$ & 
   4--11 &
   $ 0.417 (23) $ &
   $ 0.72 $ & 
   --- &
   --- & 
   --- 
\\
    $\eta' $ &
   $(1,0,0)$ & 
   5--10 &
   $ 0.471 (11) $ &
   $ 0.98 $ & 
   --- &
   --- & 
   --- 
\\
    &
   $(1,1,0)$ & 
   6--12 &
   $ 0.511 (17) $ &
   $ 1.55 $ & 
   2--4 &
   $ 0.585 (100) $ & 
   $ 0.56 $ 
\\
    &
   $(1,1,1)$ & 
   8--12 &
   $ 0.524 (77) $ &
   $ 0.66 $ & 
   --- &
   --- & 
   --- 
\\
    &
   (lat. disp.) & 
   --- &
   $ 0.392 (10) $ &
   $ 1.33 $ & 
   --- &
   --- & 
   --- 
\\
\hline
    &
   $(0,0,0)$ & 
   11--24 &
   $ 0.769 (01) $ &
   $ 0.44 $ & 
   2--7 &
   $ 0.391 (65) $ & 
   $ 0.60 $ 
\\
    $D_s $ &
   $(1,0,0)$ & 
   12--24 &
   $ 0.808 (02) $ &
   $ 0.37 $ & 
   2--7 &
   $ 0.365 (56) $ & 
   $ 0.45 $ 
\\
    &
   $(1,1,0)$ & 
   12--24 &
   $ 0.846 (03) $ &
   $ 0.24 $ & 
   2--8 &
   $ 0.360 (52) $ & 
   $ 0.27 $ 
\\
    &
   $(1,1,1)$ & 
   10--24 &
   $ 0.885 (03) $ &
   $ 0.32 $ & 
   2--8 &
   $ 0.399 (38) $ & 
   $ 0.40 $ 
\\
    &
   $(2,0,0)$** & 
   10--24 &
   $ 0.921 (04) $ &
   $ 0.24 $ & 
   2--6 &
   $ 0.407 (47) $ & 
   $ 0.26 $ 
\\
    &
   $(2,1,0)$** & 
   10--24 &
   $ 0.951 (05) $ &
   $ 0.28 $ & 
   2--6 &
   $ 0.379 (39) $ & 
   $ 0.35 $ 
\\
    &
   (lat. disp.)* & 
   --- &
   $ 0.768 (01) $ &
   $ 0.48 $ & 
   --- &
   --- & 
   --- 
\\
\end{tabular}
\hspace*{\fill} *: Not used.\\
\hspace*{\fill} **: Not included in the dispersion relation.%
\end{ruledtabular}
\end{table}

\begin{table}
\caption{
The $D_s \to \eta \ell \bar{\nu}_{\ell}$ scalar form factor $f_0(q^2)$ for Set
  A. The first column is a representative example for a given
  momentum combination. The number of equivalent
  combinations that we averaged over is given in the last column.
  The fit ranges and
  $t_{\rm sep}$ used in the simultaneous fits are also listed.  For the
  fit function, $R(t)$ in Eq.~(\protect\ref{eq:const_plus_excited}), ``${\rm 2
    exp.}+c$'' indicates that all terms are included, i.e. $c$, $A_1$ and $A_2$
  are free parameters of the fit.
}
\label{tab:f0_eta_setA}
\begin{ruledtabular}
\begin{tabular}{l|rrlccccccc}

\multicolumn{1}{c|}{$\V{p}, \V{q}, \V{k}\times L/(2\pi)$}
 & \multicolumn{1}{c}{$a^2q^2 $} & \multicolumn{1}{c}{$ f(q^2) $} & \multicolumn{1}{c}{$ t_{\rm sep}/a $ [fit range]} & fit func. & $ \chi^2/{\rm d.o.f} $ & equiv. \\
\hline
   $(0,0,0)$, $(1,0,0)$, $(-1,0,0)$ & 
    $ 0.129 (05) $ &
    $ -0.615 (19) $ & 
     8[2--6], 10[2--8], 16[9--14] & 
    $ \rm{ 2 exp.}+c$ &
    $ 1.01 $ & 
    $  6 $ 
\\
   $(1,0,0)$, $(0,0,0)$, $(1,0,0)$ & 
    $ 0.234 (06) $ &
    $ -0.659 (27) $ & 
     8[2--6], 10[2--8], 16[9--14] & 
    $ \rm{ 2 exp.}+c$ &
    $ 0.85 $ & 
    $  6 $ 
\\
   $(1,0,0)$, $(1,1,0)$, $(0,-1,0)$ & 
    $ 0.097 (06) $ &
    $ -0.592 (19) $ & 
     8[2--6], 10[2--8], 16[9--14] & 
    $ \rm{ 2 exp.}+c$ &
    $ 0.71 $ & 
    $  24 $ 
\\
   $(1,0,0)$, $(2,0,0)$, $(-1,0,0)$ & 
    $ -0.040 (06) $ &
    $ -0.544 (18) $ & 
     8[2--6], 10[2--8], 16[9--14] & 
    $ \rm{ 2 exp.}+c$ &
    $ 0.53 $ & 
    $  6 $ 
\\
   $(1,1,0)$, $(1,0,0)$, $(0,1,0)$ & 
    $ 0.202 (07) $ &
    $ -0.643 (29) $ & 
     8[2--6], 10[2--8], 16[10--14] & 
    $ \rm{ 2 exp.}+c$ &
    $ 0.72 $ & 
    $  24 $ 
\\
   $(1,1,0)$, $(1,1,1)$, $(0,0,-1)$ & 
    $ 0.065 (07) $ &
    $ -0.583 (22) $ & 
     8[2--6], 10[2--8], 16[9--14] & 
    $ \rm{ 2 exp.}+c$ &
    $ 0.65 $ & 
    $  24 $ 
\\
   $(1,1,1)$, $(1,1,0)$, $(0,0,1)$ & 
    $ 0.172 (10) $ &
    $ -0.607 (30) $ & 
     8[2--6], 10[2--8], 16[9--14] & 
    $ \rm{ 2 exp.}+c$ &
    $ 0.68 $ & 
    $  24 $ 
\\
   $(0,0,0)$, $(1,1,0)$, $(-1,-1,0)$ & 
    $ -0.023 (10) $ &
    $ -0.515 (21) $ & 
     8[3--6], 10[3--8], 16[10--14] & 
    $ \rm{ 2 exp.}+c$ &
    $ 0.74 $ & 
    $  12 $ 
\\
   $(1,0,0)$, $(0,1,0)$, $(1,-1,0)$ & 
    $ 0.073 (11) $ &
    $ -0.539 (27) $ & 
     8[3--6], 10[3--8], 16[10--14] & 
    $ \rm{ 2 exp.}+c$ &
    $ 0.55 $ & 
    $  24 $ 
\\
   $(1,0,0)$, $(1,1,1)$, $(0,-1,-1)$ & 
    $ -0.064 (11) $ &
    $ -0.481 (24) $ & 
     8[3--6], 10[3--8], 16[10--14] & 
    $ \rm{ 2 exp.}+c$ &
    $ 0.56 $ & 
    $  24 $ 
\\
   $(1,1,0)$, $(1,0,1)$, $(0,1,-1)$ & 
    $ 0.034 (12) $ &
    $ -0.526 (30) $ & 
     8[3--6], 10[3--8], 16[11--14] & 
    $ \rm{ 2 exp.}+c$ &
    $ 0.95 $ & 
    $  48 $ 
\\
   $(1,1,0)$, $(2,0,0)$, $(-1,1,0)$ & 
    $ -0.103 (12) $ &
    $ -0.450 (30) $ & 
     8[3--6], 10[3--8], 16[11--14] & 
    $ \rm{ 2 exp.}+c$ &
    $ 0.82 $ & 
    $  24 $ 
\\
\end{tabular}
\end{ruledtabular}
\end{table}

\begin{table}
\caption{
The $D_s\to \eta'\ell\bar{\nu}_{\ell}$ scalar form factor $f_0(q^2)$ for Set A,
  displayed as in Table~\protect\ref{tab:f0_eta_setA}.
  ``${\rm 1 exp.}+c$'' indicates that the parameter $A_2$ is set to zero in
the fit.}
\label{tab:f0_etap_setA}
\begin{ruledtabular}
\begin{tabular}{l|rrlccccccc}

\multicolumn{1}{c|}{$\V{p}, \V{q}, \V{k}\times L/(2\pi)$}
 & \multicolumn{1}{c}{$a^2q^2 $} & \multicolumn{1}{c}{$ f(q^2) $} & \multicolumn{1}{c}{$ t_{\rm sep}/a $ [fit range]} & fit func. & $ \chi^2/{\rm d.o.f} $ & equiv. \\
\hline
   $(0,0,0)$, $(1,0,0)$, $(-1,0,0)$ & 
    $ 0.035 (14) $ &
    $ 0.394 (44) $ & 
     8[2--6], 10[4--8], 16[10--14] & 
    $ \rm{ 2 exp.}+c$ &
    $ 1.23 $ & 
    $  6 $ 
\\
   $(1,0,0)$, $(0,0,0)$, $(1,0,0)$ & 
    $ 0.130 (15) $ &
    $ 0.395 (76) $ & 
     8[4--6], 10[4--8], 16[12--14] & 
    $ \rm{ 2 exp.}+c$ &
    $ 1.37 $ & 
    $  6 $ 
\\
   $(1,0,0)$, $(1,1,0)$, $(0,-1,0)$ & 
    $ -0.007 (15) $ &
    $ 0.357 (50) $ & 
     8[2--6], 10[4--8], 16[10--14] & 
    $ \rm{ 2 exp.}+c$ &
    $ 1.21 $ & 
    $  24 $ 
\\
   $(1,0,0)$, $(2,0,0)$, $(-1,0,0)$ & 
    $ -0.144 (15) $ &
    $ 0.356 (39) $ & 
     8[2--6], 10[4--8], 16[10--14] & 
    $ \rm{ 2 exp.}+c$ &
    $ 0.85 $ & 
    $  6 $ 
\\
   $(1,1,0)$, $(1,0,0)$, $(0,1,0)$ & 
    $ 0.089 (17) $ &
    $ 0.491 (80) $ & 
     8[2--6], 10[4--8], 16[11--12] & 
    $ \rm{ 2 exp.}+c$ &
    $ 1.15 $ & 
    $  24 $ 
\\
   $(1,1,0)$, $(1,1,1)$, $(0,0,-1)$ & 
    $ -0.048 (17) $ &
    $ 0.350 (56) $ & 
     8[2--6], 10[4--8], 16[11--13] & 
    $ \rm{ 2 exp.}+c$ &
    $ 1.25 $ & 
    $  24 $ 
\\
   $(0,0,0)$, $(1,1,0)$, $(-1,-1,0)$ & 
    $ -0.088 (17) $ &
    $ 0.375 (26) $ & 
     8[3--5], 10[4--7], 16[10--13] & 
    $ \rm{ 1 exp.}+c$ &
    $ 0.50 $ & 
    $  12 $ 
\\
   $(1,0,0)$, $(0,1,0)$, $(1,-1,0)$ & 
    $ -0.000 (20) $ &
    $ 0.416 (36) $ & 
     8[3--5], 10[5--7], 16[10--13] & 
    $ \rm{ 1 exp.}+c$ &
    $ 0.46 $ & 
    $  24 $ 
\\
   $(1,0,0)$, $(1,1,1)$, $(0,-1,-1)$ & 
    $ -0.137 (20) $ &
    $ 0.360 (29) $ & 
     8[3--5], 10[4--7], 16[10--13] & 
    $ \rm{ 1 exp.}+c$ &
    $ 0.56 $ & 
    $  24 $ 
\\
   $(1,1,0)$, $(0,0,0)$, $(1,1,0)$ & 
    $ 0.089 (22) $ &
    $ 0.510 (59) $ & 
     8[4--5], 10[4--7], 16[11--13] & 
    $ \rm{ 1 exp.}+c$ &
    $ 0.82 $ & 
    $  12 $ 
\\
   $(1,1,0)$, $(1,0,1)$, $(0,1,-1)$ & 
    $ -0.048 (22) $ &
    $ 0.441 (42) $ & 
     8[3--4], 10[4--7], 16[10--13] & 
    $ \rm{ 1 exp.}+c$ &
    $ 0.53 $ & 
    $  48 $ 
\\
   $(1,1,0)$, $(2,0,0)$, $(-1,1,0)$ & 
    $ -0.185 (22) $ &
    $ 0.353 (33) $ & 
     8[3--5], 10[4--7], 16[10--13] & 
    $ \rm{ 1 exp.}+c$ &
    $ 0.73 $ & 
    $  24 $ 
\\
   $(1,1,1)$, $(1,0,0)$, $(0,1,1)$ & 
    $ 0.043 (25) $ &
    $ 0.481 (71) $ & 
     8[4--5], 10[4--7], 16[12--13] & 
    $ \rm{ 1 exp.}+c$ &
    $ 0.62 $ & 
    $  24 $ 
\\
   $(2,0,0)$, $(1,1,0)$, $(1,-1,0)$ & 
    $ 0.001 (29) $ &
    $ 0.432 (70) $ & 
     8[4--5], 10[5--7], 16[11--13] & 
    $ \rm{ 1 exp.}+c$ &
    $ 0.88 $ & 
    $  24 $ 
\\
   $(0,0,0)$, $(1,1,1)$, $(-1,-1,-1)$ & 
    $ -0.201 (12) $ &
    $ 0.259 (41) $ & 
     8[4--5], 10[6--7], 16[11--13] & 
    $ \rm{ 1 exp.}+c$ &
    $ 1.21 $ & 
    $  8 $ 
\\
   $(1,0,0)$, $(0,1,1)$, $(1,-1,-1)$ & 
    $ -0.125 (19) $ &
    $ 0.266 (56) $ & 
     8[4--5], 10[6--7], 16[12--13] & 
    $ \rm{ 1 exp.}+c$ &
    $ 0.95 $ & 
    $  24 $ 
\\
   $(1,1,0)$, $(0,0,1)$, $(1,1,-1)$ & 
    $ -0.048 (25) $ &
    $ 0.336 (93) $ & 
     8[4--5], 10[6--7], 16[12--13] & 
    $ \rm{ 1 exp.}+c$ &
    $ 0.68 $ & 
    $  24 $ 
\\
\end{tabular}
\end{ruledtabular}
\end{table}

\begin{table}
\caption{
The $D_s\to \eta\ell\bar{\nu}_{\ell}$ scalar form factor $f_0(q^2)$ for Set S,
 displayed as in Table~\protect\ref{tab:f0_eta_setA}. 
Note that the $\chi^2/{\rm d.o.f}$ refer to uncorrelated fits.}
\label{tab:f0_eta_setS}
\begin{ruledtabular}
\begin{tabular}{l|rrlccccccc}

\multicolumn{1}{c|}{$\V{p}, \V{q}, \V{k}\times L/(2\pi)$}
 & \multicolumn{1}{c}{$a^2q^2 $} & \multicolumn{1}{c}{$ f(q^2) $} & \multicolumn{1}{c}{$ t_{\rm sep}/a $ [fit range]} & fit func. & $ \chi^2/{\rm d.o.f} $ & equiv. \\
\hline
   $(0,0,0)$, $(0,0,0)$, $(0,0,0)$ & 
    $ 0.348 (02) $ &
    $ -0.827 (14) $ & 
     16[7--10], 24[13--19] & 
    $ \rm{ 2 exp.}+c$ &
    $ 0.10 $ & 
    $  1 $ 
\\
   $(1,0,0)$, $(1,0,0)$, $(0,0,0)$ & 
    $ 0.328 (03) $ &
    $ -0.812 (17) $ & 
     16[7--10], 24[13--19] & 
    $ \rm{ 2 exp.}+c$ &
    $ 0.20 $ & 
    $  6 $ 
\\
   $(1,1,0)$, $(1,1,0)$, $(0,0,0)$ & 
    $ 0.309 (04) $ &
    $ -0.786 (25) $ & 
     16[8--11], 24[13--19] & 
    $ \rm{ 2 exp.}+c$ &
    $ 0.01 $ & 
    $  12 $ 
\\
   $(1,1,1)$, $(1,1,1)$, $(0,0,0)$ & 
    $ 0.293 (05) $ &
    $ -0.773 (29) $ & 
     16[8--10], 24[13--19] & 
    $ \rm{ 2 exp.}+c$ &
    $ 0.01 $ & 
    $  8 $ 
\\
   $(0,0,0)$, $(1,0,0)$, $(-1,0,0)$ & 
    $ 0.133 (02) $ &
    $ -0.610 (16) $ & 
     8[4--5], 10[4--7], 16[4--13], 24[13--19] & 
    $ \rm{ 2 exp.}+c$ &
    $ 0.51 $ & 
    $  6 $ 
\\
   $(1,0,0)$, $(0,0,0)$, $(1,0,0)$ & 
    $ 0.239 (02) $ &
    $ -0.658 (25) $ & 
     8[4--5], 10[4--7], 16[4--13], 24[12--19] & 
    $ \rm{ 2 exp.}+c$ &
    $ 0.20 $ & 
    $  6 $ 
\\
   $(1,0,0)$, $(1,1,0)$, $(0,-1,0)$ & 
    $ 0.102 (02) $ &
    $ -0.597 (14) $ & 
     8[4--5], 10[4--7], 16[4--13], 24[11--19] & 
    $ \rm{ 2 exp.}+c$ &
    $ 0.26 $ & 
    $  24 $ 
\\
   $(1,0,0)$, $(2,0,0)$, $(-1,0,0)$ & 
    $ -0.036 (02) $ &
    $ -0.546 (13) $ & 
     8[4--5], 10[4--7], 16[4--13], 24[15--20] & 
    $ \rm{ 2 exp.}+c$ &
    $ 0.57 $ & 
    $  6 $ 
\\
   $(1,1,0)$, $(1,0,0)$, $(0,1,0)$ & 
    $ 0.209 (03) $ &
    $ -0.640 (20) $ & 
     8[4--5], 10[4--7], 16[4--13] & 
    $ \rm{ 2 exp.}+c$ &
    $ 0.35 $ & 
    $  24 $ 
\\
   $(1,1,0)$, $(1,1,1)$, $(0,0,-1)$ & 
    $ 0.072 (03) $ &
    $ -0.582 (16) $ & 
     8[4--5], 10[4--7], 16[4--13], 24[13--19] & 
    $ \rm{ 2 exp.}+c$ &
    $ 0.20 $ & 
    $  24 $ 
\\
   $(0,0,0)$, $(1,1,0)$, $(-1,-1,0)$ & 
    $ -0.015 (03) $ &
    $ -0.550 (23) $ & 
     8[3--5], 10[3--7], 16[10--13], 24[18--21] & 
    $ \rm{ 2 exp.}+c$ &
    $ 0.07 $ & 
    $  12 $ 
\\
   $(1,0,0)$, $(0,1,0)$, $(1,-1,0)$ & 
    $ 0.083 (04) $ &
    $ -0.571 (32) $ & 
     8[3--5], 10[3--7], 16[8--13], 24[19--21] & 
    $ \rm{ 2 exp.}+c$ &
    $ 0.02 $ & 
    $  24 $ 
\\
   $(1,0,0)$, $(1,1,1)$, $(0,-1,-1)$ & 
    $ -0.054 (04) $ &
    $ -0.513 (22) $ & 
     8[3--5], 10[3--7], 16[8--13], 24[17--21] & 
    $ \rm{ 2 exp.}+c$ &
    $ 0.31 $ & 
    $  24 $ 
\\
   $(1,1,0)$, $(1,0,1)$, $(0,1,-1)$ & 
    $ 0.046 (05) $ &
    $ -0.571 (29) $ & 
     8[3--5], 10[3--7], 16[8--13], 24[17--21] & 
    $ \rm{ 2 exp.}+c$ &
    $ 0.20 $ & 
    $  48 $ 
\\
   $(1,1,0)$, $(2,0,0)$, $(-1,1,0)$ & 
    $ -0.091 (05) $ &
    $ -0.498 (21) $ & 
     8[2--6], 10[2--8], 16[2--14], 24[13--21] & 
    $ \rm{ 2 exp.}+c$ &
    $ 0.36 $ & 
    $  24 $ 
\\
   $(0,0,0)$, $(1,1,1)$, $(-1,-1,-1)$ & 
    $ -0.131 (06) $ &
    $ -0.517 (33) $ & 
     8[2--6], 10[2--8], 16[10--14], 24[17--21] & 
    $ \rm{ 2 exp.}+c$ &
    $ 0.32 $ & 
    $  8 $ 
\\
   $(1,0,0)$, $(0,1,1)$, $(1,-1,-1)$ & 
    $ -0.039 (07) $ &
    $ -0.504 (36) $ & 
     8[2--6], 10[2--8], 16[11--14] & 
    $ \rm{ 2 exp.}+c$ &
    $ 0.58 $ & 
    $  24 $ 
\\
\end{tabular}
\end{ruledtabular}
\end{table}

\begin{table}
\caption{
The $D_s\to \eta'\ell\bar{\nu}_{\ell}$ scalar form factor $f_0(q^2)$ for Set S,
displayed as in Table~\protect\ref{tab:f0_eta_setA}.
``${\rm 1 exp.}+c$'' indicates that the parameter $A_2$ is set to zero in
the fit.}
\label{tab:f0_etap_setS}
\begin{ruledtabular}

\begin{tabular}{l|rrlccccccc}

\multicolumn{1}{c|}{$\V{p}, \V{q}, \V{k}\times L/(2\pi)$}
 & \multicolumn{1}{c}{$a^2q^2 $} & \multicolumn{1}{c}{$ f(q^2) $} & \multicolumn{1}{c}{$ t_{\rm sep}/a $ [fit range]} & fit func. & $ \chi^2/{\rm d.o.f} $ & equiv. \\
\hline
   $(1,0,0)$, $(1,0,0)$, $(0,0,0)$ & 
    $ 0.086 (15) $ &
    $ 0.523 (45) $ & 
     8[2--6], 10[4--8], 16[12--14] & 
    $ \rm{ 1 exp.}+c$ &
    $ 1.12 $ & 
    $  6 $ 
\\
   $(1,1,0)$, $(1,1,0)$, $(0,0,0)$ & 
    $ 0.049 (16) $ &
    $ 0.495 (54) $ & 
     8[2--6], 10[4--8], 16[12--14] & 
    $ \rm{ 1 exp.}+c$ &
    $ 1.00 $ & 
    $  12 $ 
\\
   $(0,0,0)$, $(1,0,0)$, $(-1,0,0)$ & 
    $ 0.019 (06) $ &
    $ 0.430 (24) $ & 
     8[2--5], 10[3--7], 16[10--13] & 
    $ \rm{ 1 exp.}+c$ &
    $ 0.39 $ & 
    $  6 $ 
\\
   $(1,0,0)$, $(0,0,0)$, $(1,0,0)$ & 
    $ 0.113 (07) $ &
    $ 0.464 (35) $ & 
     8[2--5], 10[3--7], 16[10--13] & 
    $ \rm{ 1 exp.}+c$ &
    $ 0.48 $ & 
    $  6 $ 
\\
   $(1,0,0)$, $(1,1,0)$, $(0,-1,0)$ & 
    $ -0.024 (07) $ &
    $ 0.417 (27) $ & 
     8[2--5], 10[3--7], 16[9--13] & 
    $ \rm{ 1 exp.}+c$ &
    $ 0.25 $ & 
    $  24 $ 
\\
   $(1,0,0)$, $(2,0,0)$, $(-1,0,0)$ & 
    $ -0.161 (07) $ &
    $ 0.371 (27) $ & 
     8[2--5], 10[3--7], 16[9--13] & 
    $ \rm{ 1 exp.}+c$ &
    $ 0.50 $ & 
    $  6 $ 
\\
   $(1,1,0)$, $(1,0,0)$, $(0,1,0)$ & 
    $ 0.071 (07) $ &
    $ 0.451 (42) $ & 
     8[2--5], 10[3--7], 16[11--13] & 
    $ \rm{ 1 exp.}+c$ &
    $ 0.52 $ & 
    $  24 $ 
\\
   $(1,1,0)$, $(1,1,1)$, $(0,0,-1)$ & 
    $ -0.066 (07) $ &
    $ 0.415 (33) $ & 
     8[2--5], 10[3--7], 16[10--13] & 
    $ \rm{ 1 exp.}+c$ &
    $ 0.26 $ & 
    $  24 $ 
\\
   $(1,1,1)$, $(1,1,0)$, $(0,0,1)$ & 
    $ 0.033 (08) $ &
    $ 0.468 (44) $ & 
     8[2--5], 10[3--7], 16[12--13] & 
    $ \rm{ 1 exp.}+c$ &
    $ 0.38 $ & 
    $  24 $ 
\\
   $(0,0,0)$, $(1,1,0)$, $(-1,-1,0)$ & 
    $ -0.069 (08) $ &
    $ 0.403 (32) $ & 
     8[2--6], 10[4--8], 16[9--14] & 
    $ \rm{ 2 exp.}+c$ &
    $ 0.98 $ & 
    $  12 $ 
\\
   $(1,0,0)$, $(0,1,0)$, $(1,-1,0)$ & 
    $ 0.022 (10) $ &
    $ 0.427 (45) $ & 
     8[2--6], 10[4--8], 16[10--14] & 
    $ \rm{ 2 exp.}+c$ &
    $ 0.52 $ & 
    $  24 $ 
\\
   $(1,0,0)$, $(1,1,1)$, $(0,-1,-1)$ & 
    $ -0.115 (10) $ &
    $ 0.355 (37) $ & 
     8[2--6], 10[4--8], 16[11--14] & 
    $ \rm{ 2 exp.}+c$ &
    $ 0.58 $ & 
    $  24 $ 
\\
   $(1,1,0)$, $(1,0,1)$, $(0,1,-1)$ & 
    $ -0.022 (11) $ &
    $ 0.400 (54) $ & 
     8[2--6], 10[4--8], 16[11--14] & 
    $ \rm{ 2 exp.}+c$ &
    $ 0.36 $ & 
    $  48 $ 
\\
   $(1,1,0)$, $(2,0,0)$, $(-1,1,0)$ & 
    $ -0.159 (11) $ &
    $ 0.373 (45) $ & 
     8[2--6], 10[4--8], 16[10--14] & 
    $ \rm{ 2 exp.}+c$ &
    $ 0.54 $ & 
    $  24 $ 
\\
   $(0,0,0)$, $(1,1,1)$, $(-1,-1,-1)$ & 
    $ -0.147 (33) $ &
    $ 0.369 (34) $ & 
     8[2--5], 10[5--7], 16[10--13] & 
    $ \rm{ 1 exp.}+c$ &
    $ 0.82 $ & 
    $  8 $ 
\\
   $(1,0,0)$, $(0,1,1)$, $(1,-1,-1)$ & 
    $ -0.058 (39) $ &
    $ 0.342 (45) $ & 
     8[3--5], 10[5--7], 16[10--13] & 
    $ \rm{ 1 exp.}+c$ &
    $ 0.87 $ & 
    $  24 $ 
\\
\end{tabular}
\end{ruledtabular}
\end{table}
\newpage~~
\newpage~~
\newpage~~
\newpage
\bibliography{references}
\end{document}